\documentclass[a4paper,12pt]{article}
\pdfoutput=1 
\usepackage{jheppub}
\usepackage{subcaption}
\usepackage{amscd,amsmath,amssymb,amsthm,comment,dsfont,graphicx,psfrag}
\usepackage[dvipsnames]{xcolor}
\usepackage{lipsum}  
\usepackage{float}
\usepackage{slashed}
\usepackage[percent]{overpic}  

\hypersetup{bookmarksdepth=3, bookmarksnumbered=true, bookmarksopen=false} 


\newcommand{\beq}{\begin{equation}}
\newcommand{\eeq}{\end{equation}}
\newcommand{\beqr}{\begin{displaymath}}
\newcommand{\eeqr}{\end{displaymath}}
\newcommand{\beqa}{\begin{eqnarray}}
\newcommand{\eeqa}{\end{eqnarray}}
\newcommand{\beqar}{\begin{eqnarray*}}
\newcommand{\eeqar}{\end{eqnarray*}}

\newcommand{\half}{\ensuremath{\frac{1}{2}}}
\newcommand{\cF}{\ensuremath{\mathfrak{F}}}

\preprint{}

\title{\boldmath Discrete basis parameterization\\ for the  Gauge Theory Bootstrap}

\author{Rafael Córdoba}
\affiliation{Laboratoire de Physique de l’Ecole Normale Supérieure }
\affiliation{ ENS-Université PSL,
CNRS, Sorbonne Université, Université Paris Cité,\\ F-75005 Paris, France}
\emailAdd{rafael.cordoba@ens.fr}

\abstract{We implement the Gauge Theory Bootstrap (GTB) framework, initiated by He and Kruczenski in \cite{He:2023lyy, He:2024nwd}, using a discrete basis parametrization of the   2-to-2 pion scattering \( S \)-matrix, the spectral densities and the form factors.  This approach enables a refined analysis of the convergence of the GTB and drastically reduced computational time —from approximately half an hour in \cite{He:2024nwd} to under one minute in ours. The discrete basis  also facilitates  the evaluation of the \( S \)-matrix across the complex sheet resulting, in additional of the dominant \( \rho(770) \) and \( f_2(1270) \) resonances identified in \cite{He:2024nwd}, in the extraction of the \( \sigma \) meson pole located far from the real axis. 
}
\begin{document}

\phantomsection 
\pdfbookmark[1]{Abstract}{Abstract} 

\maketitle

\flushbottom
\newpage
\section{Introduction}

The modern S-matrix bootstrap program \cite{Paulos_2017,Paulos_20172,paulos2017smatrixbootstrapiiihigher,kruczenski2022snowmass} revives the old idea \cite{Eden:1966dnq} that physical  amplitudes are required to satisfy consistency constraints --the analyticity, crossing, and unitarity (ACU) constraints-- by  making those constraints computationally concrete.  These developments allowed  novel, non-perturbative explorations of scattering amplitudes including, in the context of pion scattering, bounds on low energy parameters \cite{Guerrieri:2018uew,10.21468/SciPostPhys.10.5.122,10.21468/SciPostPhys.9.5.081,Guerrieri:2020bto,Fernandez2023,Albert:2022oes,Albert2024,AlbertJan}.  A  improvement of the S-matrix bootstrap was further introduced in \cite{Karateev:2019ymz}, where the bootstrap philosophy was extended beyond amplitudes to include additional observables, such as form factors and spectral densities, thereby encoding further physical information into the bootstrap setup.

Building on these ideas, the recently proposed Gauge Theory Bootstrap (GTB) \cite{He:2023lyy} explores the strongly coupled dynamics of QCD by combining the S-matrix bootstrap principles (the ACU constraints) with both, IR information coming from chiral symmetry breaking and, UV input from perturbative QCD. The IR information is, for instance, incorporated by requiring the S-matrix to reproduce, at low energies, the tree-level amplitude \cite{Weinberg:1966kf,GASSER1984142}, while the UV input is encoded through the SVZ Finite Energy Sum Rules (FESR) for spectral functions   \cite{Shifman:1978bx,Shifman:1978by,SHIFMAN1979519} and the asymptotic behavior of form factors \cite{Brodsky:1989pv,Tong_2022,TONG2021136751,He:2024nwd}. Enforcing the amplitude to satisfy the ACU constraints, the IR behavior, and the UV behavior on spectral functions and form factors defines the  GTB constraints, whose solutions can be found with the help of semi-definite programming.

 To numerically implement the GTB framework,  \cite{He:2023lyy,He:2024nwd} parametrized the GTB observables, i.e. the amplitude, form factors and  spectral densities, using an interpolation point method.   Imposing the GTB constraints at these interpolation points, for  finitely many constraints --as this often yields good convergence \cite{paulos2017smatrixbootstrapiiihigher}--, results in a finite-dimensional convex optimization problem, whose solutions  yield  the  2-to-2 pion scattering amplitudes, from factors and spectral densities.
 The outcome of this procedure is promising, offering first-principles insights into the non-perturbative dynamics of QCD and demonstrating remarkable agreement with both experimental and phenomenological QCD data \cite{He:2024nwd,He:2023lyy,he2025gaugetheorybootstrappredicting}.

Although the GTB yielded encouraging results, it was desirable to improve the computational cost: the complexity scaled significantly with the number of interpolation points $M$, limiting an efficient exploration of the observables.
In this work, we present an alternative implementation of the GTB using the discrete basis  introduced in \cite{He:2021eqn}, for the $S$-matrix, form factors, and spectral densities. This parametrization expresses each observable  as a Taylor series in the discrete basis variable, very much in line with the ideas of \cite{paulos2017smatrixbootstrapiiihigher}\footnote{Such discrete basis has be used previously in \cite{Chen:2021pgx} for studying the $\phi^4$ theory by combining the form factor bootstrap of \cite{Karateev:2019ymz}
with Hamiltonian truncation. }. To implement the GTB numerically, we thus consider the observable's  partial sums, until an order where good convergence is achieved, reducing the problem of determining the observables to that of finding a finite set of Taylor coefficients. Imposing the GTB constraints on these coefficients also yields a finite-dimensional convex optimization problem, whose  solutions are  our bootstrapped observables.

This  approach provides two main advantages. First, it serves as a check of the results obtained by He and Kruczenski \cite{He:2024nwd}, thereby further corroborating the GTB framework. Second, it enables a  study of the convergence properties of the bootstrap procedure and helps identify the minimal number of coefficients needed --to reliably discard higher order Taylor coefficients--, as well as the minimal number of constraints we impose to those.
The result of this procedure allows us to drastically reduced computational time --from approximately
half an hour in \cite{He:2023lyy,He:2024nwd} to under one minute in ours-- while achieving comparable results.

The organization of this paper is as follows. In Section \ref{sec:partialwaves}, we briefly review the GTB framework, followed in Section \ref{sec:BGTpar} by the discrete parametrization of the $S$-matrix, form factors, and spectral densities. Section \ref{sec:num} details the numerical implementation and shows the resulting 2d space of allowed amplitudes consistent with the ACU constraints alone,  with ACU supplemented by IR information, and with the full set of GTB constraints. In  section \ref{sec:convnum}, we  present graphical convergences of the GTB on the  numerical parameters.
Section \ref{sec:res} then presents our main results, including the first six partial waves obtained from the discrete parametrization. We also plot the  pole structure of the bootstrapped amplitudes on the $S0$, $P1$ and $D0$ channel. Finally, Section \ref{sec:out} summarizes the work done.

 \paragraph{Note }
 While this paper is being completed,  new ingredients on the GTB where introduced in \cite{he2025gaugetheorybootstrappredicting}. Although here we follow the initial setup of \cite{He:2023lyy,He:2024nwd}, it  would be good to consider the new ingredients of \cite{he2025gaugetheorybootstrappredicting} for future works and improvements.

\section{Summary of the Gauge Theory Bootstrap}
\label{sec:partialwaves}
The theory we consider is an $SU(3)$  gauge theory in 4d with matter content consisting of two quark flavours --i.e. QCD--. We assume that the theory undergoes spontaneous chiral symmetry breaking and confinement. The low-energy physics, thus, is governed by the three pions $\pi_a$, which transform under the isospin symmetry $SU(2)_V$ and appear as the Goldstone bosons of the symmetry breaking. We consider the lightest scattering process in the theory: the pion-pion scattering,
\begin{equation}
 \pi_a(p_1)+\pi_b(p_2)\to\pi_c(p_3)+\pi_d(p_4).
\label{amplitudepion}
\end{equation}
Letting $P=p_1+p_2$  be the in-coming momentum and $s=P^2$, the physical observables we are going to compute are 
the $S$-matrix elements in each partial wave channel\footnote{Here we use the  angular momentum and isospin partial wave decomposition,
\begin{equation}
f^I_\ell(s) = \frac{1}{4} \int_{-1}^1 d\cos\theta P_{\ell}(\cos\theta) T^I(s,t),
\label{fdef}
\end{equation}
where  $P_\ell$ are the Legendre polynomials, $T^I$ is the $T$-Matrix ($S=1+iT$) with defined isospin, and   $t=\frac{(s-4)(1-\cos \theta)}{2}$.}
\begin{equation}
S^I_\ell(s)=1+i f_\ell^I(s), \quad s \in \mathbb{R}{\ge 4},
\label{eq:funSf1}
\end{equation}
 the form factors
\beqa
F_\ell^I(s) := \,_{out}\langle \pi(p_2)\pi(p_1) | \mathcal{O}_\ell^I(0) | 0 \rangle,
\eeqa 
and the spectral densities 
\begin{equation}
\text{Spec}^I_\ell(s)  := \int \frac{d^4x}{(2\pi)^4} e^{iP \cdot x} \langle 0 | \mathcal{O}_\ell^I(x)^\dagger \mathcal{O}_\ell^I(0) | 0 \rangle,
\end{equation}
where  the local operators $\mathcal{O}_\ell^I$ are taken from \cite{He:2024nwd}
\begin{equation}
    \mathcal{O}^0_0 =  m_q(\bar{u}u+\bar{d}d),\quad  \mathcal{O}^1_{1} = \half(\bar{u}\gamma^\mu u -\bar{d} \gamma^\mu d),\quad  \mathcal{O}^0_2 = T^{++},
    \label{eq:currents}
\end{equation}
$m_q$ is the quark mass (here we take $m_q=(m_u+m_d)/2$), $u$ and $d$ are the standard up and down spinors, $T^{\mu\nu}$ is the energy-momentum tensor, and we have used the coordinates $x^\pm=\frac{1}{\sqrt{2}}(x\pm iy)$. In order to bootstrap these observables, we impose the  
Analyticity, Crossing Symmetry and Unitarity (ACU) constraints together with the UV and IR  information defined as follows:
\subsection*{Analyticity, Crossing Symmetry and Unitarity constraints}
In addition to the $S-$matrix unitarity  constraint,
\begin{equation}
    |S^I_\ell(s)|\leq 1,\quad \, s\in \mathbb{R}_{\geq 4}
    \label{spartial}
\end{equation}
where, due to Bose statistics, $\ell \in \text{Mod}_2(I) + 2\mathbb{N}$, 
    we impose the positive semi-definite condition \cite{Karateev:2019ymz}:
\begin{equation}
B :=
\left( \begin{array}{ccc}
1 & S_\ell^I(s) & \mathcal{F}_\ell^{I}(s) \\
{S}_\ell^{*I}(s) & 1 & \mathcal{F}_\ell^{*I}(s) \\
\mathcal{F}_\ell^{*I}(s) & \mathcal{F}_\ell^I(s) & \text{Spec}_\ell^I(s)
\end{array} \right) \succeq 0.
\label{bound1}
\end{equation}
Here $\mathcal{F}_\ell^{I}$ are the form factors in the  momentum $P$ and angular momentum $\ell$ \footnote{
The relation between the form factors $\cF_\ell^I$ and $F_\ell^I$, for the $S0$, $P1$ and $D0$ channels --up to Kronecker delta functions projecting into the basis--  are given by \cite{He:2024nwd}:
	\begin{align}
	\cF_0^0(s) &=\frac{\sqrt{3}}{16 \pi ^{5/2}}\left(\frac{s-4}{s}\right)^{1/4} F_0^0(s) ,\label{FFS0scale}\\
    \cF_1^1(s) &= \frac{\sqrt{s}}{8 \sqrt{6} \pi ^{5/2}}\left(\frac{s-4}{s}\right)^{3/4} F_1^1(s) ,\label{FFP1scale}\\
\cF^0_2(s) &= \frac{s}{16 \sqrt{10} \pi ^{5/2} } \left( \frac{s-4}{s} \right)^{5/4} F^0_2(s) \label{FFD0scale}.
	\end{align}}. Notice that eq. \eqref{bound1}  couple together  the $S-$Matrix elements, form factors and, spectral densities. 
    
    Maximal analyticity and crossing symmetry  of the $S$-matrix elements, on the other hand, is  made manifest by parametrizing it in terms of a  scattering amplitude  $A(s,t,u)$ --where   $s$, $t$, and $u$ are the Mandelstam variables--as follows:
\begin{align}
T_{ab,cd} &= A(s,t,u) \delta_{ab} \delta_{cd} + A(t,s,u)\delta_{ac} \delta_{bd} + A(u,t,s) \delta_{ad} \delta_{bc}
\label{eqh}
\end{align}
with $A(s,t,u)$ defined in terms of the Mandelstam representation \cite{He:2021eqn}:
\begin{equation}
    \begin{aligned}A(s,t,u)= T_0&+\frac1\pi\int_4^\infty dx\frac{\sigma^{(1)}(x)}{x-s}+\frac1\pi\int_4^\infty dx\:\sigma^{(2)}(x)\left[\frac1{x-t}+\frac1{x-u}\right]\\&\hspace{1.2cm}+\frac1{\pi^2}\int_4^\infty dx\int_4^\infty dy\frac{\rho^{(1)}(x,y)}{x-s}\left[\frac1{y-t}+\frac1{y-u}\right]\\&\hspace{2.8cm}+\frac1{\pi^2}\int_4^\infty dx\int_4^\infty dy\frac{\rho^{(2)}(x,y)}{(x-t)(y-u)},\end{aligned}
    \label{spec}
\end{equation}
where $T_0$ is a constant,  $\sigma^{(\alpha)}(x)$ and $\rho^{(\alpha)}(x,y)$ for $\alpha = 1,2$ are spectral functions parametrizing the amplitude $A(s,t,u)$ and, the function $\rho^{(2)}(x,y)$ is required to satisfy $\rho^{(2)}(x,y) = \rho^{(2)}(y,x)$.  Similarly, analyticity of the form factor is manifest via the dispersion relation \cite{He:2023lyy}:
\begin{equation}
F^I_\ell(s) = 1 + \frac{1}{\pi} \int_4^\infty dx \left( \frac{1}{x - s} - \frac{1}{x} \right) \Im F^I_\ell(x),
\label{Fdisp}
\end{equation}
where we have imposed the normalization condition $F^I_\ell(0) = 1$ and we have set $m_\pi = 1$ in all of these expressions. analyticity of two point function of local operators $\mathcal{O}_\ell^I$ implies its associated  spectral functions  $\text{Spec}_\ell^I(s)$ has support on the physical region $s\in [4,\infty).$

\subsection*{IR information}
Using chiral perturbation theory, the tree-level amplitude $A(s,t,u)$,
at low energies,  takes the form
\cite{Weinberg:1966kf, Donoghue_Golowich_Holstein_2014}:
\begin{equation}
  A^\chi(s,t,u)=\frac{s-m_\pi^2}{8\pi^2f_\pi^2},
  \label{ampchi}
\end{equation}
which allows us to determine the $S-$matrix element partial waves $f_\ell^I(s)$, defined by eq. \eqref{fdef}, as
\begin{equation}
    f_{0,\mathrm{tree}}^0(s)=\frac{2}{\pi}\frac{2s-m_\pi^2}{32\pi f_\pi^2},\quad f_{1,\mathrm{tree}}^1(s)=\frac{2}{\pi}\frac{s-4m_\pi^2}{96\pi f_\pi^2},\quad f_{0,\mathrm{tree}}^2(s)=\frac{2}{\pi}\frac{2m_\pi^2-s}{32\pi f_\pi^2},
    \label{eq:fchiral}
\end{equation}
where $f_\pi$ is the pion decay constant.
 Following \cite{He:2023lyy,He:2024nwd}, we match the tree-level amplitudes at  low energies by imposing the constraints:
\begin{equation}
||f^{I,\mathrm{boot}}_{0}(s_j)-R^\chi_{I1}(s_j)f^{1,\mathrm{boot}}_{1}(s_j)||\leq\varepsilon_{CSB},\quad  R_{I1}^\chi(s)=\frac{f_{0,\mathrm{tree}}^I(s)}{f_{1,\mathrm{tree}}^1(s)}.
\label{CSBconst}
\end{equation}
 Here, $f^{I,\mathrm{boot}}_\ell(s_j)$ denotes the partial waves $f_\ell^I(s)$ computed from the parametrization \eqref{spec} and $I=0,2$. The constraints \eqref{CSBconst} are imposed at sample points $\tilde s_j \in (0, 4)$, thereby incorporating IR behavior.

Although in this paper we only consider the IR behavior of the scattering amplitude, as pointed out in \cite{he2025gaugetheorybootstrappredicting}, the set of operators \eqref{eq:currents} can be identified in the free pion theory from which one   can include further constraints on the low energy behavior of the spectral functions. These new ingredients of the GTB turns out to be quite powerful \cite{he2025gaugetheorybootstrappredicting}, whose study we leave for future work.

\subsection*{UV information}
  UV information can be encoded using  Finite Energy Sum Rule (FESR) \cite{Shifman:1978bx,Shifman:1978by,SHIFMAN1979519} and  form factor's asymptotic behavior \cite{Brodsky:1989pv} for $s > s_0$, where $s_0$ is an energy scale at which pQCD is consider to be valid. Leading order behavior of both of those can be found from pQCD, whose  values   read \cite{He:2024nwd}:
    	\begin{subequations}
	\beqa
	\frac{1}{s_0^{n+2}}\int_4^{s_0} \text{Spec}^0_0(x) x^n dx &\simeq&  \frac{3 \left(m_d^2+m_u^2\right)}{64 \pi ^5 (n+2)} \left(1+\alpha_S\frac{13}{3 \pi }\right),\ n\ge 0 \label{S0SR}, \\
	\frac{1}{s_0^{n+2}}\int_4^{s_0} \text{Spec}^1_1(x) x^n dx &\simeq&  \frac{1}{64 \pi ^5 (n+2)}\left(1+\frac{\alpha_S}{\pi }\right),\ \ n\ge -1\label{P1SR}, \\
	\frac{1}{s_0^{n+3}} \int_4^{s_0}\!\text{Spec}^0_2(x) x^n dx&\simeq&\frac{1}{64 \pi ^5 (n+3)}\left(\frac{11}{10}-\frac{17 \alpha_S}{18 \pi }\right), \ n\ge -2 \label{D0SR},
	\eeqa
    \label{sumrulestheory}
	\end{subequations}	
	\begin{subequations}
	\beqa
    |F^0_0(s\to \infty)| &\simeq &   4\pi\alpha_S  f_\pi^2\frac{\left(m_d^2+m_u^2\right)}{s}\log s, \ \ \label{S0asymp} \\
    |F^1_1(s\to \infty)| &\simeq &   16\pi\frac{\alpha_S  f_\pi^2}{s}, \ \  \label{P1asymp}\\
    | F^0_2(s\to \infty) | &\simeq& \frac{48\pi\alpha_S  f_\pi^2}{s},\; \label{D0asymp}
	\eeqa
    \label{asymFFtheory}
	\end{subequations}	
where $m_d$ and $m_u$ are the masses of the two quark flavours, $f_\pi$ is the pion decay constant, and $\alpha_S$ is the QCD coupling.
  Following \cite{He:2023lyy,He:2024nwd}, these high-energy predictions 
   translates into constraints on the bootstrap through the inequalities: 
	\begin{subequations}
	\beqa
     \left|\int_{4}^{s_{0}}\text{Spec}^{boot}(x)x^{n}dx-\int_{4}^{s_{0}}\text{Spec}(x)x^{n}dx\right|&\leq&\varepsilon_{SVZ},
  \label{SVZconst}\\
\left| F^{\text{boot}}(s > s_0) \right| &\leq &\varepsilon_{\text{FF}},
\label{FF}
	\eeqa
    \label{uvconst}
	\end{subequations}	
where we have suppressed isospin and angular momentum indices. Here 
$\text{Spec}^{\text{boot}}$ and $F^{boot}$ are  functions we want to determine. Because of eq. \eqref{Fdisp},  one  further parametrize the  $F^{boot}$  function in terms of the imaginary part of the form factor, $\Im F^{I,boot}_\ell$.

All together, the GTB framework encodes  the data in the following functions:
\begin{equation}
 \left\{T_0,\,\sigma^{(\alpha)}(x), \, \rho^{(\alpha)}(x,y) ,\ \text{Im}{F}^{I,boot}_\ell(x),\, \text{Spec}_\ell^{I,boot}(x)\right\}.
\label{paremeters}
\end{equation}
 Functions \eqref{paremeters}  are thus found --bootstrapped-- by requiring them to follow the ACU constraints \eqref{spartial} and \eqref{bound1} and, the IR and UV constraints from eq.s \eqref{CSBconst} and \eqref{uvconst}.

\section{Discrete Basis Parametrization}
\label{sec:BGTpar}
In order to bootstrap the functions \eqref{paremeters}, we implement a different parametrization from that of \cite{He:2023lyy,He:2024nwd} (see Section 8 of \cite{He:2024nwd} for details); we parametrize the observables in terms of a discrete basis defined as follows. Consider the  map \cite{paulos2017smatrixbootstrapiiihigher} \begin{equation}
     z(s)=\frac{\sqrt{4-\nu_0}-\sqrt{4-s}}{\sqrt{4-\nu_0}+\sqrt{4-s}}
     \label{mapzz}
\end{equation}
that sends the first complex sheet of the Mandelstam variable $s$ into the unit disk. We choose $\nu_0=0$ so that $s=0$ is mapped to the center of the disk. Under this map, the physical region $s\in [4,\infty)$ gets identified with the upper-half circle by 
$$z(s+i \epsilon)=e^{i \varphi},$$ where the $i\varepsilon$ prescription ensures we are working in the first complex sheet.
We then consider the discrete basis of functions $\{\phi_n(x)\}_{n\in \mathbb{N}_{\geq 0}}$ defined by \cite{He:2021eqn}:
\begin{equation}
\phi_n(x)  = \sin(n\varphi), \quad x(\varphi) = \frac{4}{\cos^2\frac{\varphi}{2}}, \quad n \geq 1.
\label{discretebasisp}
\end{equation}
Using this basis, the functions parameterizing the scattering amplitude \eqref{spec}  can be expanded as
\begin{equation}
    \begin{aligned}\rho^{(\alpha)}(x,y)&=\sum_{n_1,n_2=1}^\infty \rho^{(\alpha)}_{n_1n_2}\phi_{n_1}(x)\phi_{n_2}(y), \quad \quad \sigma^{(\alpha)}(x)&=\sum_{n=1}^\infty \sigma_n^{(\alpha)}\phi_n(x),\end{aligned}
    \label{expansionbasis}
\end{equation}
where, due to $\rho^{(2)}(x,y)=\rho^{(2)}(y,x)$  symmetry, we require $\rho^{(2)}_{n m}=\rho^{(2)}_{mn }$. 
Writing the $S-$matrix elements in the form of
\begin{equation}
S^I_\ell(s) = 1 + i\pi \sqrt{\frac{s - 4}{s}} f^I_\ell(s), 
\label{eq:funSf}
\end{equation}
we can find the discrete basis parametrization of  $f_\ell^I(s)$ to be 
 \begin{equation}
  f_\ell^I(s)=\left\{T_0A_\ell^I+\sum_{n=1}^{\infty}\left[A_{\ell n}^I(s)\sigma_n^{(1)}+B_{\ell n}^I(s)\sigma_n^{(2)}\right]\right.+\sum_{n,m=1}^{\infty}\left[A_{\ell,nm}^I(s)\rho_{nm}^{(1)}
+B_{\ell,nm}^I(s)\rho_{nm}^{(2)}\right]\biggr\},
\label{funcoeff}
 \end{equation}
where the functions $A(s)$ and $B(s)$ can be found by plugging-in the discrete basis expansion \eqref{expansionbasis} on the Mandelstam representation \eqref{spec} and taking the $S-$matrix  partial waves decomposition. Explicit expressions are \cite{He:2021eqn} ($N=3$):
$$
\begin{aligned}
A_{\ell}^0 & =2(N+2) \delta_{\ell 0}, & A_{\ell}^1 & =0, & A_{\ell}^2 & =4 \delta_{\ell 0}, \\
A_{\ell, n}^0 & =2 N \delta_{\ell 0} \Phi_n+2 \hat{\Phi}_{n \ell}, & A_{\ell, n}^1 & =2 \hat{\Phi}_{n \ell},  &A_{\ell, n}^2&=2 \hat{\Phi}_{n \ell}, \\
B_{\ell, n}^0 & =4 \delta_{\ell 0} \Phi_n+2(N+1) \hat{\Phi}_{n \ell}, &  B_{\ell, n}^1&=-2 \hat{\Phi}_{n \ell}, &B_{\ell, n}^2&=4 \delta_{\ell 0} \Phi_n+2 \hat{\Phi}_{n \ell},\\
A_{\ell, n m}^0 & =2 N \Phi_n \hat{\Phi}_{m \ell}+2 \Phi_m \hat{\Phi}_{n \ell}+2 \tilde{\Phi}_{n m, \ell}, &
A_{\ell, n m}^1 & =2 \Phi_m \hat{\Phi}_{n \ell}+2 \tilde{\Phi}_{n m, \ell}, &
A_{\ell, n m}^2 & =2 \Phi_m \hat{\Phi}_{n \ell}+2 \tilde{\Phi}_{n m, \ell}, \\
B_{\ell, n m}^0 & =\Phi_n \hat{\Phi}_{m \ell}+\Phi_m \hat{\Phi}_{n \ell}+N \tilde{\Phi}_{n m, \ell}, &
B_{\ell, n m}^1 & =-\Phi_n \hat{\Phi}_{m \ell}-\Phi_m \hat{\Phi}_{n \ell}, &
B_{\ell, n m}^2 & =\Phi_n \hat{\Phi}_{m \ell}+\Phi_m \hat{\Phi}_{n \ell},
\end{aligned}
$$
where $\Phi_n$, $\hat{\Phi}_{m \ell}$ and $\tilde{\Phi}_{n m, \ell}$
are defined as 
    \begin{subequations}
    \beqa
    \Phi_n(s) &=& \frac{1}{\pi} \int_4^{\infty} dx, \frac{\phi_n(x)}{x - s} = z^n(s) - (-1)^n,
\label{phiint}\\
      \hat{\Phi}_{n, \ell}(s)  &=&\int_{-1}^{+1} d \mu P_{\ell}(\mu) \Phi_n(t),\\
\tilde{\Phi}_{n m, \ell}(s)  &=&\int_{-1}^{+1} d \mu P_{\ell}(\mu) \Phi_n(t) \Phi_m(u),
    \label{eq:intbasis}
\eeqa
\end{subequations}
with  $z(x)$ as in eq. \eqref{mapzz}.

\paragraph{Form Factor and Spectral Density  Parametrization.}
Using the discrete basis of functions \eqref{discretebasisp}, we parametrize the imaginary part of the form factor and the spectral densities as\footnote{
Note that with the new ingredients of \cite{he2025gaugetheorybootstrappredicting}, the low energy behavior can be explicitly parametrized into the basis expansion (c.f. eqs. B.20-B.22 in \cite{he2025gaugetheorybootstrappredicting}). }
\begin{align}
\Im F_\ell^I(s)=\sum_{n=1}^{\infty} ImF_{\ell,n}^I \phi_n(s),\\
\text{Spec}_\ell^I(x)=\sum_{n=1}^{\infty}\text{Spec}_{\ell,n}^I \phi_n(x),\label{specpar}
\end{align}
 where $ImF_{\ell,n}^I$ and $\text{Spec}_{\ell,n}^I$ are constant coefficients. Using the dispersion relation \eqref{Fdisp}, together with the integral \eqref{phiint},
 one finds   the form factor discrete basis parametrization:
\begin{equation}
     F_\ell^I(s)=1+\sum_{n=1}^{\infty}  ImF_{\ell,n}^I z^n(s).
\label{paramFF}
\end{equation}

All together, the GTB  parameters \eqref{paremeters} are completely characterized by  coefficients
\begin{align}
  {\bf v}:=\{T_0,\,\sigma^{(\alpha)}_n,\,\rho^{(\alpha)}_{nm},\, \text{Spec}_{\ell,n}^I\, , \mathrm{Im}F^I_{n,\ell} \}_{\alpha=1,2}^{n,m=1,2,..},
\label{eq:GTBfullparameters}
\end{align}
with $\rho^{(2)}_{nm} = \rho^{(2)}_{mn}$ and quantum numbers $(I,\ell)\in \{(0,0),(1,1),(0,2)\}$ corresponding to the $S0$, $P1$ and $D0$ form factor/spectral densities  we consider here.

\section{Numerical Implementation}
\label{sec:num}
To  implement  the bootstrap numerically, we require the unitarity  constraints \eqref{spartial} and \eqref{bound1} to be satisfied only at finitely many values. For this, we consider the physical energies  \begin{equation}
    \{s_i= x(\varphi_i)\}_{i=1,...,M} \subset [4, \infty),
    \label{energies}\end{equation}
    where $M$ is the number of energies we consider, $x(\varphi)$ is the map described in  \eqref{discretebasisp},      and $\varphi_i$ are  uniformly distributed angles given by
\begin{equation}
\varphi_i = \frac{\pi}{M}\left(i - \frac{1}{2}\right),\quad i = 1, 2, \ldots, M.
\label{genM}
\end{equation}
We also consider the partial sums of the $S-$matrix elements, form factors and spectral densities basis expansion --mainly   \eqref{funcoeff}, \eqref{specpar} and \eqref{paramFF}--, until a maximum value $n_{\text{max}}$, chosen such that partial sums converges with the desired precision. 

 The --now finitely many--GTB coefficients \eqref{eq:GTBfullparameters} that satisfy the GTB constraints --namely, eq.s \eqref{spartial}, \eqref{bound1}, \eqref{CSBconst}, and \eqref{uvconst}-- at the physical energies $\{s_i\}_{i=1}^M$, and for each partial wave $\ell \leq \ell_{\text{max}}$, yield the bootstrapped $S$-matrix, spectral functions, and form factors. Since these constraints define a finite-dimensional convex optimization problem\footnote{In general, regularizing the norm of the double spectral density $\rho(x, y)$ is necessary. This stems from the fact that highly oscillatory contributions to $\rho(x, y)$ have negligible impact on the physical amplitude. Imposing an upper bound on the norm suppresses such oscillations; see \cite{He:2021eqn} for further details.}, such coefficients can be found, for instance, using semi-definite programming solvers \cite{mosek,boyd2004convex,10.1007/978-1-84800-155-8_7,1573950400904311936}.  
 
  Since it is instructive to show how each constraint supresses portions of the allowed space of amplitudes, as done in \cite{He:2023lyy}, we proceed to show the resulting 2d space of allowed amplitudes consistent with the ACU constraints alone,  with ACU supplemented by IR information, and with the full set of GTB constraints.
\subsection{Analyticity, Crossing Symmetry and Unitarity Bootstrap}
\label{ACUboot}

At this level, the amplitude is characterized by the coefficients
\begin{equation}
   {\bf w}:=\{T_0,\,\sigma^{(\alpha)}_n,\,\rho^{(\alpha)}_{nm}\}_{\alpha=1,2}^{n,m=1,...,n_{max}},
\label{eq:ACUw}
\end{equation}
where $\rho^{(2)}$ is required to be symmetric and, $n_{max}$ is the order at which we consider the partial in the discrete basis expansion \eqref{funcoeff}.

Using the Mandelstam representation of the amplitude $A(s,t,u)$, crossing symmetry and analyticity are manifest. Unitarity, i.e.  eq. \eqref{spartial}, is then imposed only at the finitely many energies $\{s_i\}_{i=1}^M$ of eq. \eqref{energies} and for a finite number of partial waves $(I,\ell)$, where $I\in \{0, 1, 2\}$ and $\ell \leq \ell_{\text{max}}$. This results on a finite dimensional, convex problem, of $ 1 + \frac{5 n_{\text{max}} \left(3 n_{\text{max}} + 1\right)}{2}$ variables subject to $ M \cdot \ell_{\text{max}} \cdot 3$ constraints. 

\begin{figure}[h]
    \begin{center}
     \includegraphics[width=1\linewidth]{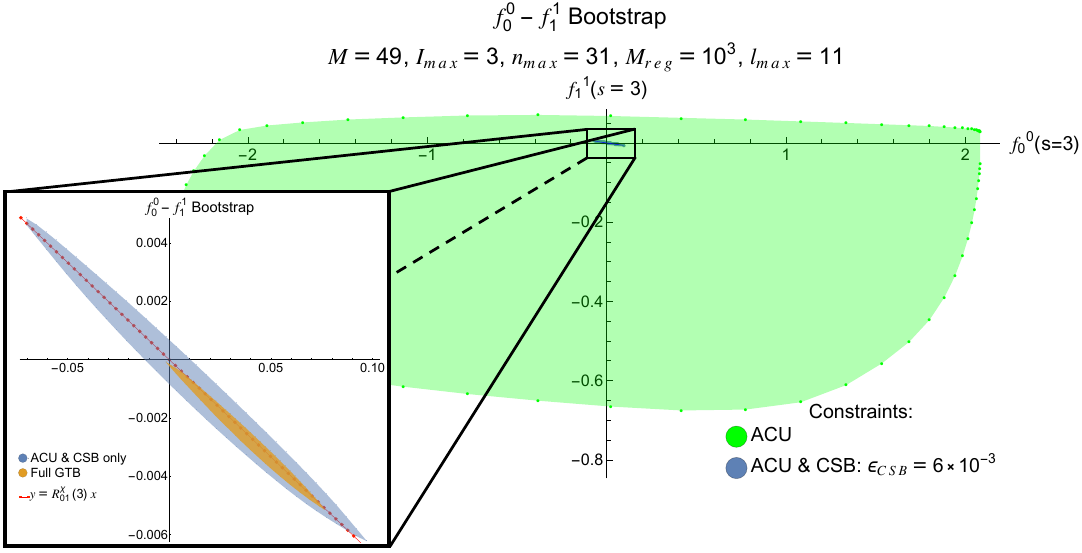}
    \caption{Gauge Theory Bootstrap: The region in green represents the  space of allowed amplitudes under the constraints of ACU. Zooming-in, in blue we observe the space of allowed amplitudes constrained by both ACU and CSB constraints (IR information). Incorporating the full GTB constraints yields the GTB-allowed space shown in orange. Within this region, the  relation between $f_0^0(3)$ and $f_1^1(3)$, predicted by tree-level chiral perturbation theory, is shown by a red dashed line.
    \label{fig:3GTB}}
  \end{center}
  \end{figure}

To compare with the results of \cite{He:2023lyy}, we consider the linear functional  $f_\ell^I(s)$ defined in eq. \eqref{funcoeff}, whose  coefficients $A(s)$ and $B(s)$'s are evaluated to 20-digit numerical precision using 
\texttt{Mathematica}. Finding the extremal values of $f_0^0(3)$ within the range of the extremal values of $f_1^1(3)$, subject to the constraint of  \eqref{spartial},  corresponds to finding the 
boundary of the ACU-allowed amplitude space on the $f_0^0(3)-f_1^1(3)$ plane. For instance, with $M = 49$, $\ell_{max}=11$ and $n_{max}=31$, the convex optimization yields the ACU-allowed space shown in green in Figure \ref{fig:3GTB}.

    Each point within the ACU-allowed  corresponds to an amplitude satisfying the ACU constraints. For boundary points, the convex optimization gives the explicit values of the ACU parameters \eqref{eq:ACUw} where constraints are saturated i.e. where $|S_\ell^I(s_i)|=1$. Substituting these solutions into eq. \eqref{funcoeff} yields the bootstrapped partial waves $f_\ell^I(s)$ for each boundary point.

\subsection{IR information and Full Gauge Theory Bootstrap}

To incorporate the IR information from Chiral Symmetry Breaking, in addition to the ACU constraints and following \cite{He:2023lyy,He:2024nwd}, we impose the relation of eq. \eqref{CSBconst} at low-energy points $\tilde{s}_i \in \{1/2,1,3/2,2\}$.

The inclusion of this IR data further constrains the amplitude space obtained from the ACU bootstrap, in agreement with the findings of \cite{He:2023lyy}. Figure \ref{fig:3GTB} illustrates this effect. The green region corresponds to the set of amplitudes satisfying only the ACU constraints. Once the CSB constraints from eq. \eqref{CSBconst} are imposed, the allowed region is significantly reduced, as shown in blue. Zooming-in on the ACU-CSB constrained space, we observe that the set of allowed amplitudes—now satisfying both ACU and CSB conditions—is tightly bounded around the relation predicted by three-level chiral perturbation theory:
\begin{equation}
y = R_{01}^\chi(3) x,
\label{relationcsb}
\end{equation}
where $ R_{01}^\chi(3)$ is the partial wave ratio  of eq. \eqref{CSBconst}. This is shown in the figure as a red dashed  line, emphasizing consistency between the bootstrap results and the effective field theory.

Having incorporated both the ACU and IR constraints we now complete the Gauge Theory Bootstrap by imposing the full  constraints; it remains to include the UV information in the form of form factor asymptotics and the FESR  of eq.s \eqref{uvconst} and \eqref{bound1}.
To evaluate the theoretical estimates appearing in eq.s \eqref{sumrulestheory} and \eqref{asymFFtheory}, which are required as input for eq. \eqref{uvconst}, we use the following physical input parameters:
\begin{align*}
s_0=(2\mathrm{GeV})^2,\quad \alpha_S(2GeV)=0.3145887440,
\quad\left(\begin{matrix}
    m_u\\
    m_d
\end{matrix}\right)_{2GeV}= \left(\begin{matrix}
    3.6 MeV\\
    6.5 MeV
\end{matrix}\right).
\end{align*}

Implementing these additional UV constraints on top of the ACU and CSB setup, shown in blue in Figure \ref{fig:3GTB}, results in a further reduction of the allowed amplitude space. This is visualized in Figure \ref{fig:3GTB} on orange, consistent with the findings of \cite{He:2023lyy}.
From the figure, we observe that incorporating the UV information --via the from factor asymptotics and sum rules-- further suppress portions of the ACU-CSB allowed space and push the allowed amplitudes toward the lower-right  region. The graph reflects the nontrivial impact of UV information on low-energy scattering data. The  information the bootstrapped amplitudes carry is discussed in Section \ref{sec:res}.

 All convex optimization and numerical computations were carried out using \texttt{Mathematica} and \texttt{MatLab}. The relevant scripts and data are available in the repository:
\href{https://github.com/rafaf991/Bootstrapping-Gauge-Theories.git}{\texttt{Bootstrapping Gauge Theories}}.
\subsection{Unitarity Improved Solutions}
\label{unitimprov}
Running the GTB  yields the bootstrapped coefficients:
\begin{align}
   \{T_0^{\text{GTB}},\,\sigma^{(\alpha),{\text{GTB}}}_n,\,\rho^{(\alpha),{\text{GTB}}}_{n,m},\,\text{Spec}_{n,\ell}^{{\text{GTB}},I},\,\mathrm{Im}F_{n,\ell}^{\text{GTB}} \}_{\alpha=1,2}^{n,m=1,...,n_{max}},
\label{functinalsFF3}
\end{align}
for each boundary point in the two-dimensional space of allowed amplitudes (Orange region in Figure \ref{fig:3GTB}). 
In particular, each solution  \eqref{functinalsFF3}  corresponds to an amplitude where unitarity \eqref{bound1} is saturated and consequently where the spectral densities are dominated by two-pion intermediate states \cite{He:2023lyy}. 

Numerically, however, the saturation of unitarity \eqref{bound1} is not fully achieved, and an additional numerical reinforcement  is required.  Following \cite{He:2024nwd, He_2018}, we implement a  unitarization reinforcement as follows. Consider a solution  \eqref{functinalsFF3} satisfying the GTB constraints.  Starting from its underlying amplitude --which is now a known function--, we  define $ h_\ell^I(s)|_{old}$ by\footnote{Here one can also  implement the 'Watsonian unitarization' where both amplitude and form factor's information is stored in $h_\ell^I(s)|_{old}$. This is done by making a different choice of $h_\ell^I(s)|_{old}$, as shown in  \cite{he2025gaugetheorybootstrappredicting}.} 
 \begin{equation}
     S^I_\ell(s) =1+ih_\ell^I(s)\Big|_{old}.
 \end{equation}
Associated to this amplitude we consider an unitarity-improved $h_\ell^I(s,{\bf w})|_{new} $ (whose parameters ${\bf w}$  of eq. \eqref{eq:ACUw} are sought to be found)   close to the initial $ h_\ell^I(s)|_{old}$ i.e. 
$h_\ell^I(s)|_{new} = h_\ell^I(s)|_{old}+ \varepsilon.$ 
Linearizing the unitarity  constraint of Equation \eqref{spartial} by plugging the solution $h_\ell^I(s)|_{old}$ gives, in terms of $h_\ell^I$,
\begin{equation}
    \hat{\mathcal{F}}_\ell^I(s,{\bf w}) =\text{Re}  h_\ell^I(s)\Big|_{old} \text{Re}h_\ell^I(s,{\bf w})\Big|_{new}+\text{Im}h_\ell^I(s)\Big|_{old}\text{Im}h_\ell^I(s,{\bf w})\Big|_{new}-\text{Im}h_\ell^I(s,{\bf w})\Big|_{new}\leq 0.
\end{equation}
 Since  improvement of saturation requires the functional $\hat{\mathcal{F}}_\ell^I$ to be as big as possible, to take into account all the partial waves, at all physical energies, we maximize the functional 
 \begin{equation}
     \begin{aligned}\mathcal{F}=\sum_{i=1}^M \sum_{\ell,I} \hat{\mathcal{F}}_\ell^I(s_i,{\bf w})\end{aligned}
 \end{equation}
 subject to the GTB constraints. Solving the convex problem yields the unitary-improved $S_\ell^I|_{new}(s)$. One can iterate the process again until a convergence is found. 
Such point is the solution we take as our bootstrapped $S$-matrix. 
\newpage
\section{Convergence Plots and GTB Numerical Dependences}
\label{sec:convnum}
 In the GTB all the relevant information is encoded in the vector of coefficients ${\bf v}(M,\ell_{\max},n_{\max})$, defined in eq. \eqref{eq:GTBfullparameters}, and where we have emphasized its dependence on the numerical parameters $M,\,\ell_{\max}$ and, $n_{\max}$. 
As seen in Section \ref{ACUboot}, the complexity of the resulting optimization problem scales quadratically with $n_{\text{max}}$ --as $L=1 + \frac{5 n_{\text{max}} \left(3 n_{\text{max}} + 1\right)}{2}$ is the number of variables--, and linearly with both $M$ and $\ell_{\text{max}}$, which determine the number of constraints. Therefore, before analyzing the physical information contained in the bootstrapped amplitudes, we analyze the convergence behavior of these numerical parameters as increasing them leads to a rapid growth in computational cost.

\subsection{$M,\,\ell_{max} $ and $n_{max}$ Dependence}
\label{sec:dependences}
\begin{figure}[H]
\centering
\begin{subfigure}[b]{0.9\linewidth}
\includegraphics[height=0.3\textheight,trim={0cm 0 0 1.3cm},clip]{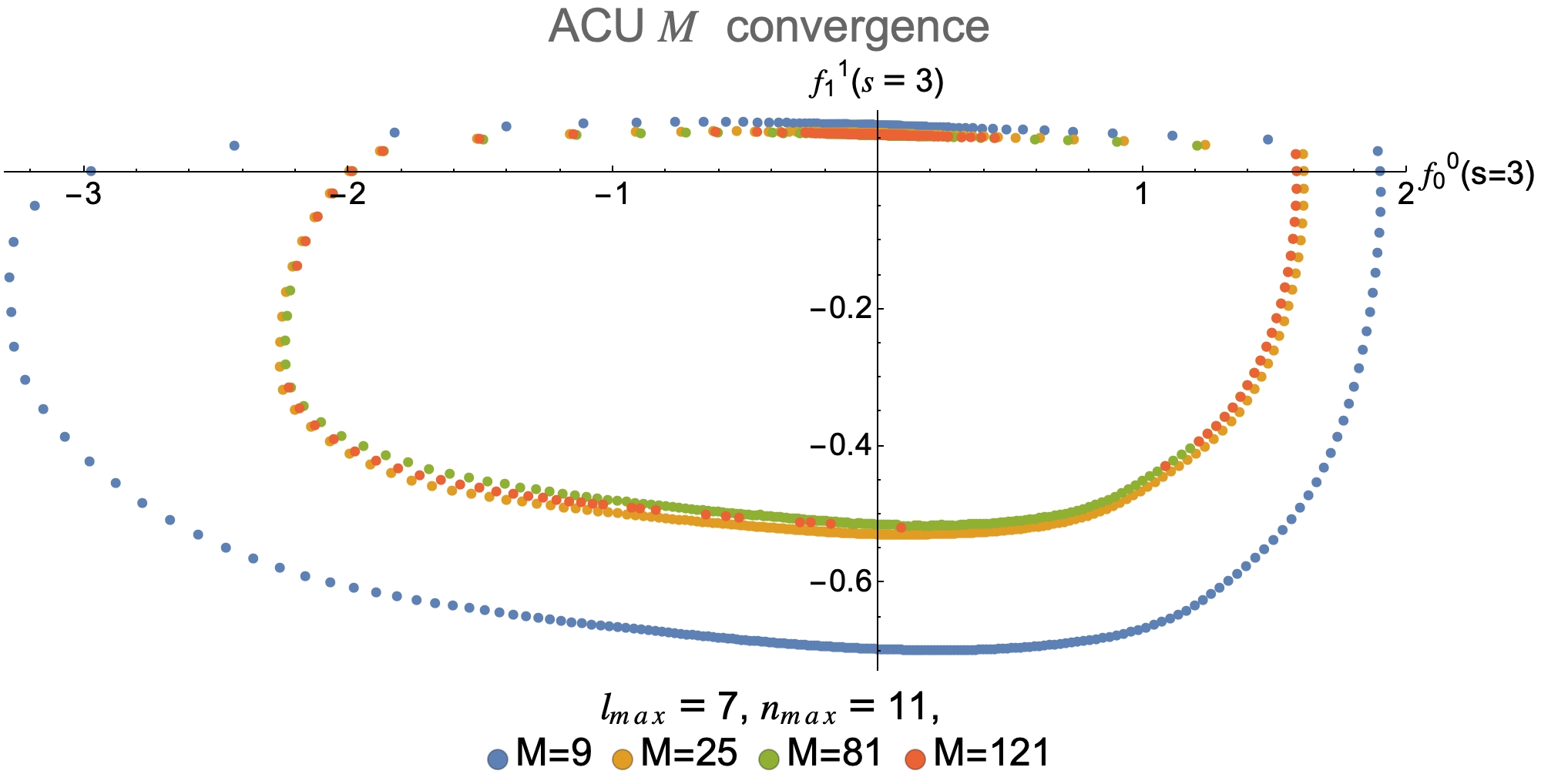}
\caption{ $M$\label{fig:ACUMconvergence}}
\end{subfigure}
\begin{subfigure}[b]{0.9\linewidth}
\includegraphics[height=0.3\textheight,trim={0cm 0 0 1.3cm},clip]{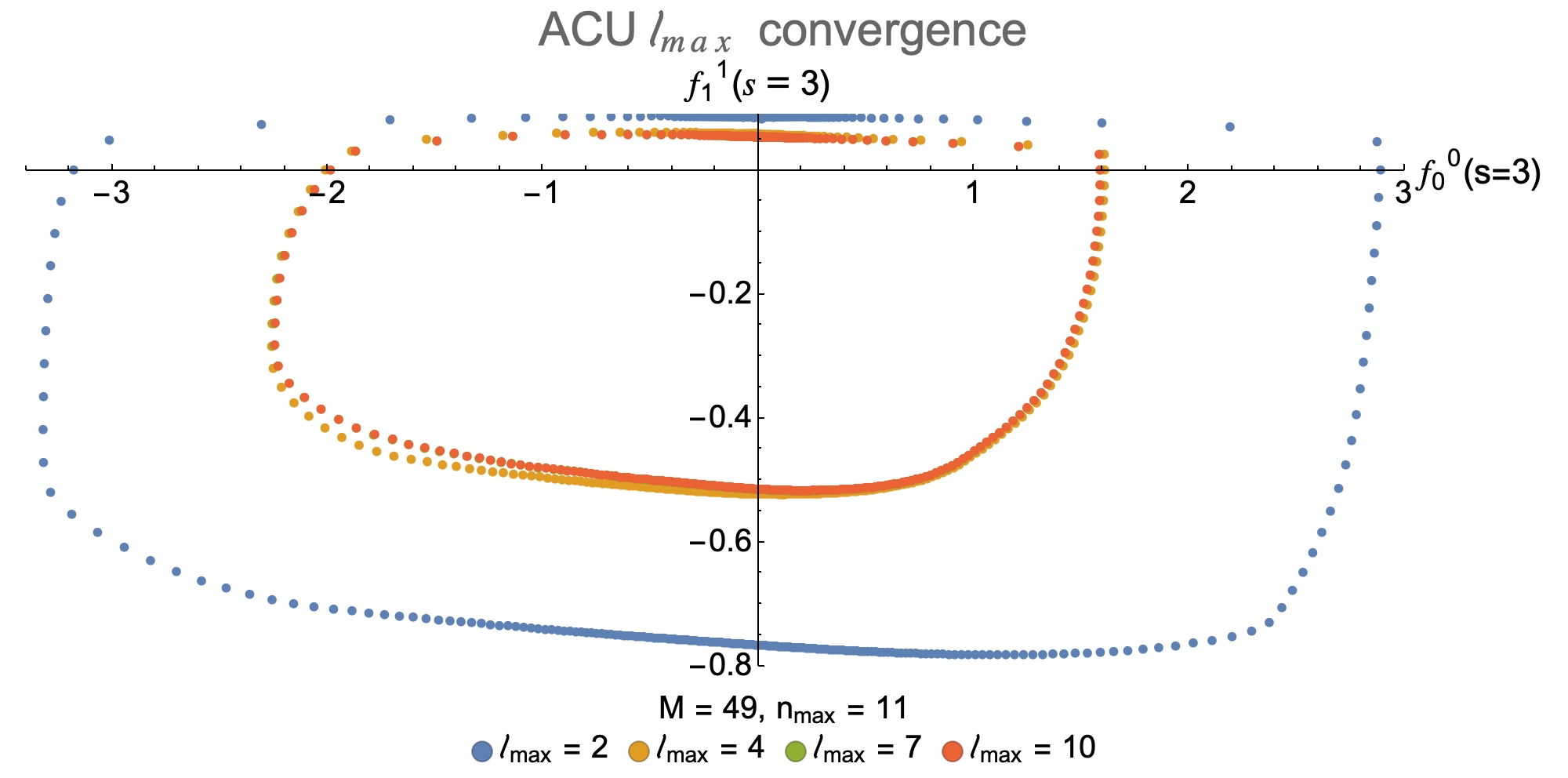}
\caption{$\ell_{\text{max}}$\label{fig:ACUlmaxconvergence}}
\end{subfigure}

\begin{subfigure}[b]{0.9\linewidth}
\includegraphics[height=0.3\textheight,trim={0cm 0 0 1.3cm},clip]{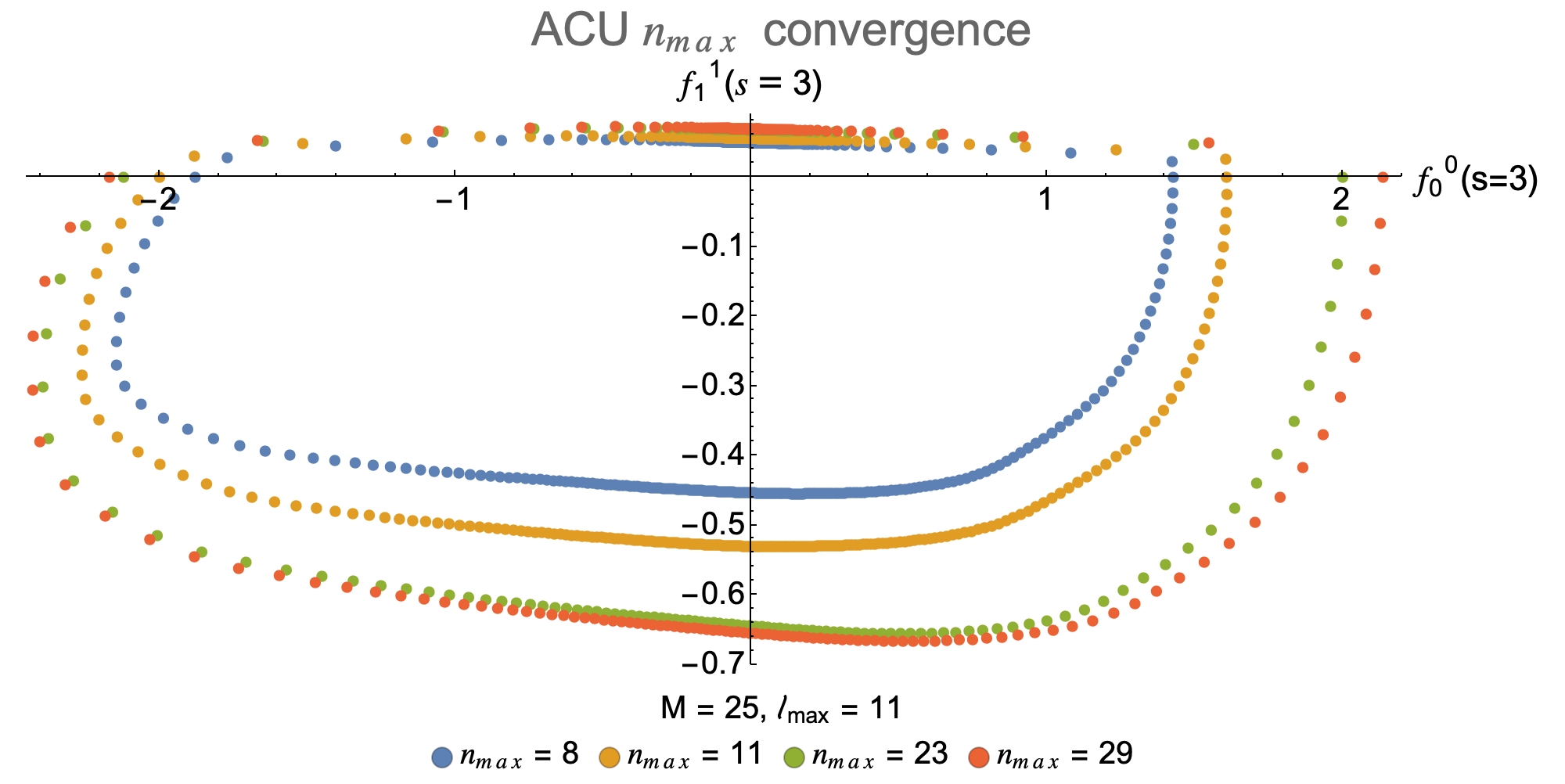}
\caption{$n_{\text{max}}$\label{fig:ACUnmaxconvergence}}
\end{subfigure}

\caption{Dependence of the allowed amplitude space (ACU) on the parameters $M$, $\ell_{\text{max}}$ and $n_{max}$.\label{fig:convergence}}
\end{figure}

In Figure \ref{fig:convergence} we present graphical convergences illustrating how the discrete basis bootstrap --imposing only the ACU constraints-- depends on the $M$, $\ell_{max}$ and $n_{max}$ parameters, varying one parameter at a time while keeping the others fixed.

In Figure \ref{fig:ACUMconvergence}, we observe that the ACU-allowed space of amplitudes stabilizes for $M\geq 25$.
At $M = 121$, some points are missing near the lower-right kink due to weak unfeasibility in the optimization. This was resolved by increasing the regularization parameter to $M_{\text{Reg}} > 1000$. This issue does not affect the full GTB analysis: the FESR and form factor asymptotics effectively constrain the high-frequency oscillations in the amplitude making the choice of regularization superfluous.

Convergence in $\ell_{\text{max}}$ is shown in Figure \ref{fig:ACUlmaxconvergence} for fixed $M = 49$ and $n_{\text{max}} = 11$. We observe that the space of allowed amplitudes, projected onto $f_0^0(3)$ and $f_1^1(3)$, stabilizes after including just four nontrivial partial waves\footnote{Here we refer to the lower-right quadrant of the allowed region, where the physical value of  $(f_0^0(3),f_1^1(3))$ is expected to be.}.  
Similarly, convergence in $n_{\text{max}}$ is shown in Figure \ref{fig:ACUnmaxconvergence} for fixed $\ell_{\text{max}} = 7$ and $M = 25$.   We observe  good convergence around  $n_{max}\sim 29$.

\begin{figure}[H]

\begin{subfigure}[b]{0.49\linewidth}
\hspace{0.1cm}
\begin{overpic}[width=\linewidth,trim={2.5cm 5.13cm 2.5cm 0cm},clip]{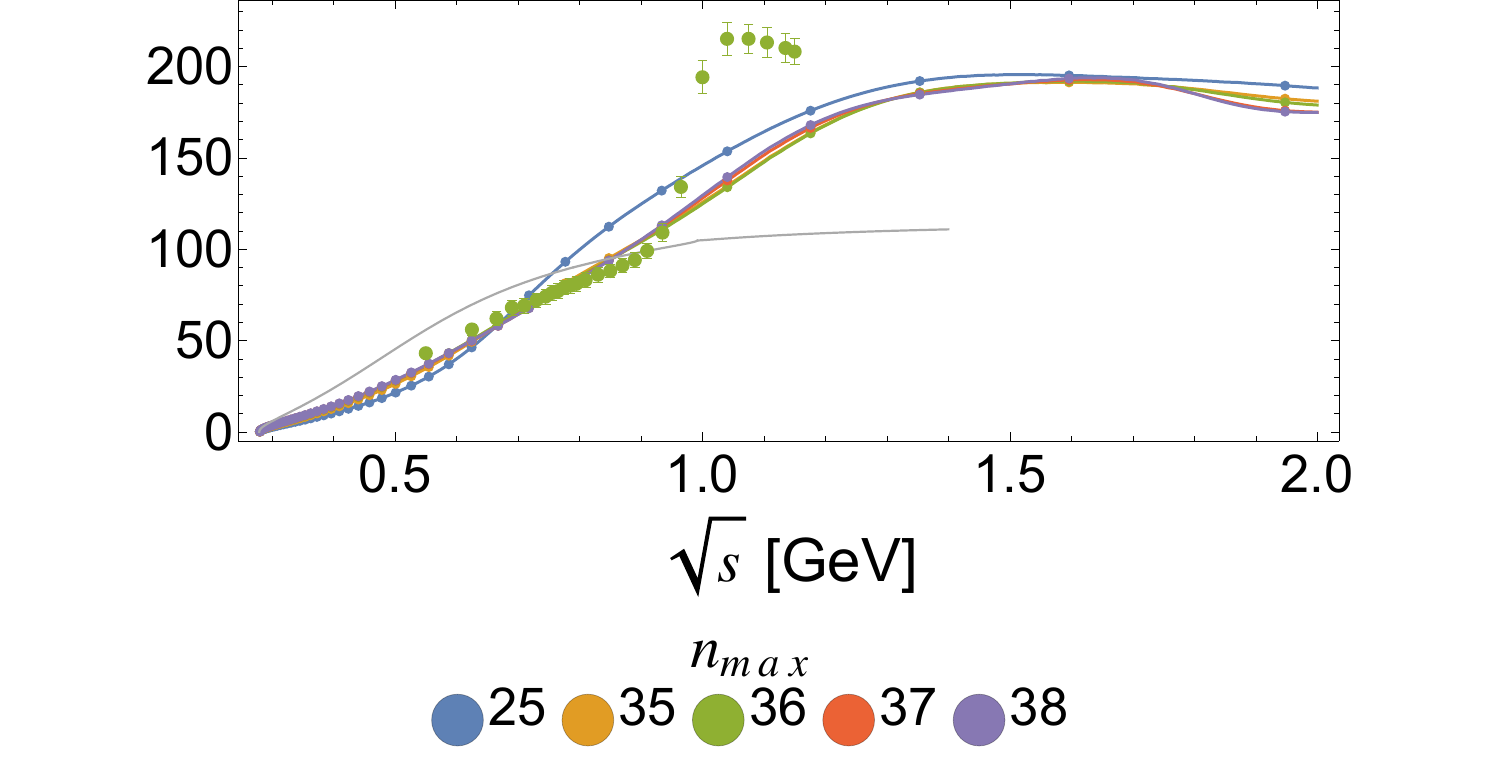}
 \put(15,30){\Large $\delta_0^0$}
\end{overpic}
\begin{overpic}[width=1.017\linewidth,trim={4.93cm 0.2cm 7.34cm 0cm},clip]{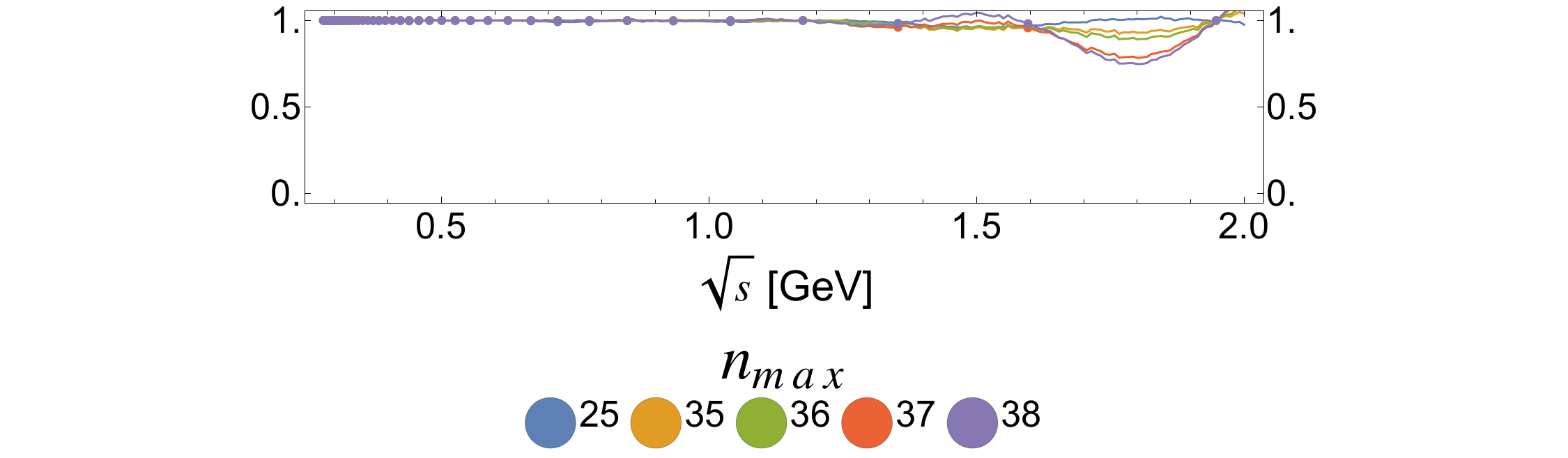}
\put(15,30){\Large $\eta_0^0$}
\end{overpic}
\caption{$S0$ channel.}
\label{fig:anmax}
\end{subfigure}
\begin{subfigure}[b]{0.49\linewidth}
\hspace{0.1cm}\begin{overpic}[width=\linewidth,trim={2.5cm 4.64cm 2.5cm 0cm},clip]{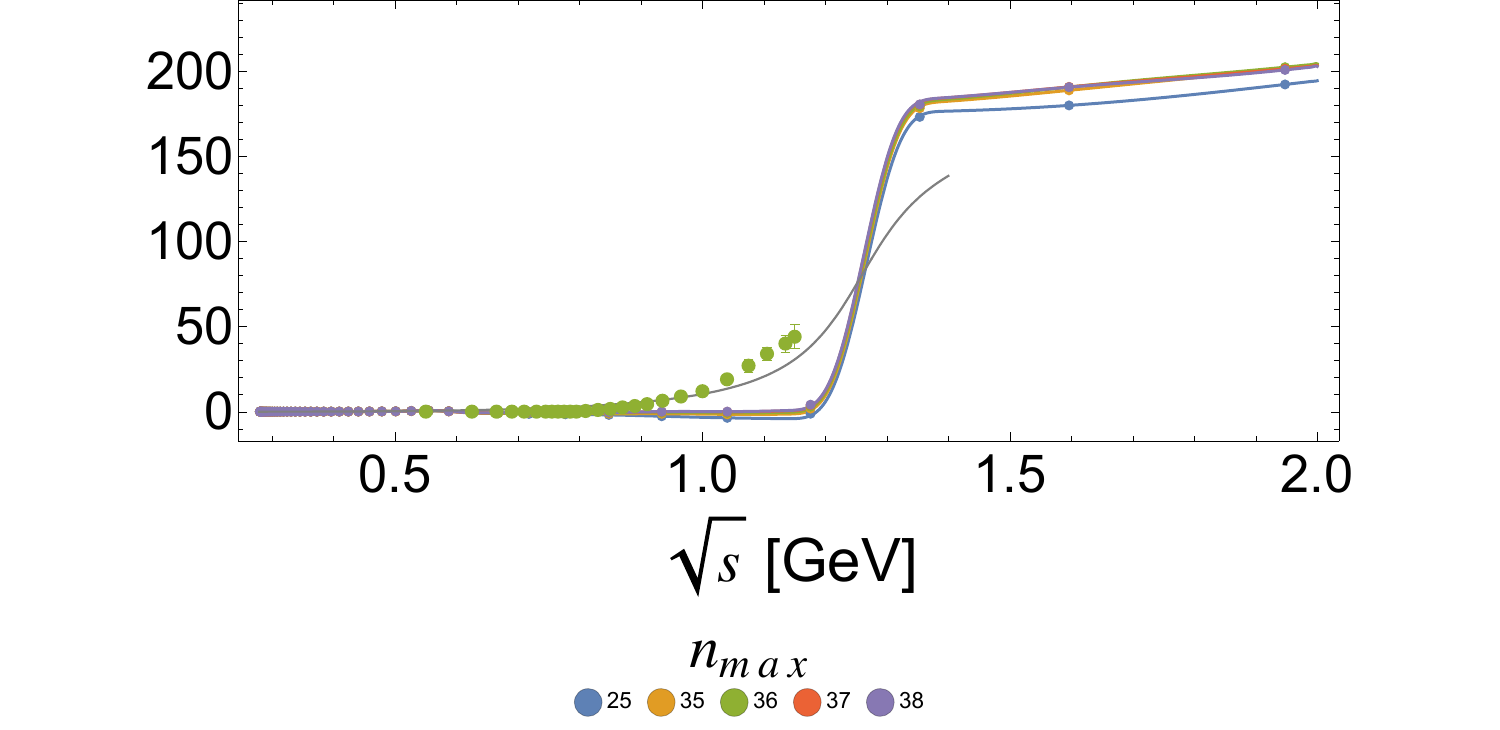}
\put(15,30){\Large $\delta_2^0$}
\end{overpic}
\begin{overpic}[width=1.006\linewidth,trim={5.2cm 0.2cm 7.34cm 0cm},clip]{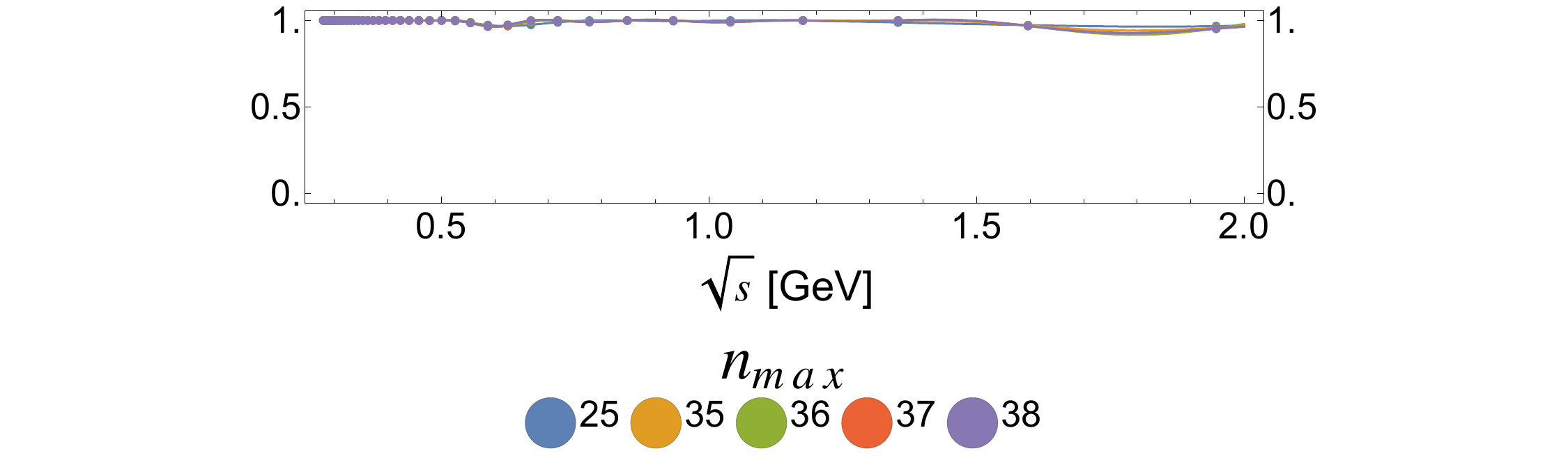}
\put(15,30){\Large $\eta_2^0$}
\end{overpic}
\caption{$D0$ channel.}
\label{fig:bnmax}
\end{subfigure}

\centering
\begin{subfigure}[b]{0.49\linewidth}
\begin{center}
    
{\begin{overpic}
    [width=1.014\linewidth,trim={2.55cm 4.92cm 2.72cm 0cm},clip]{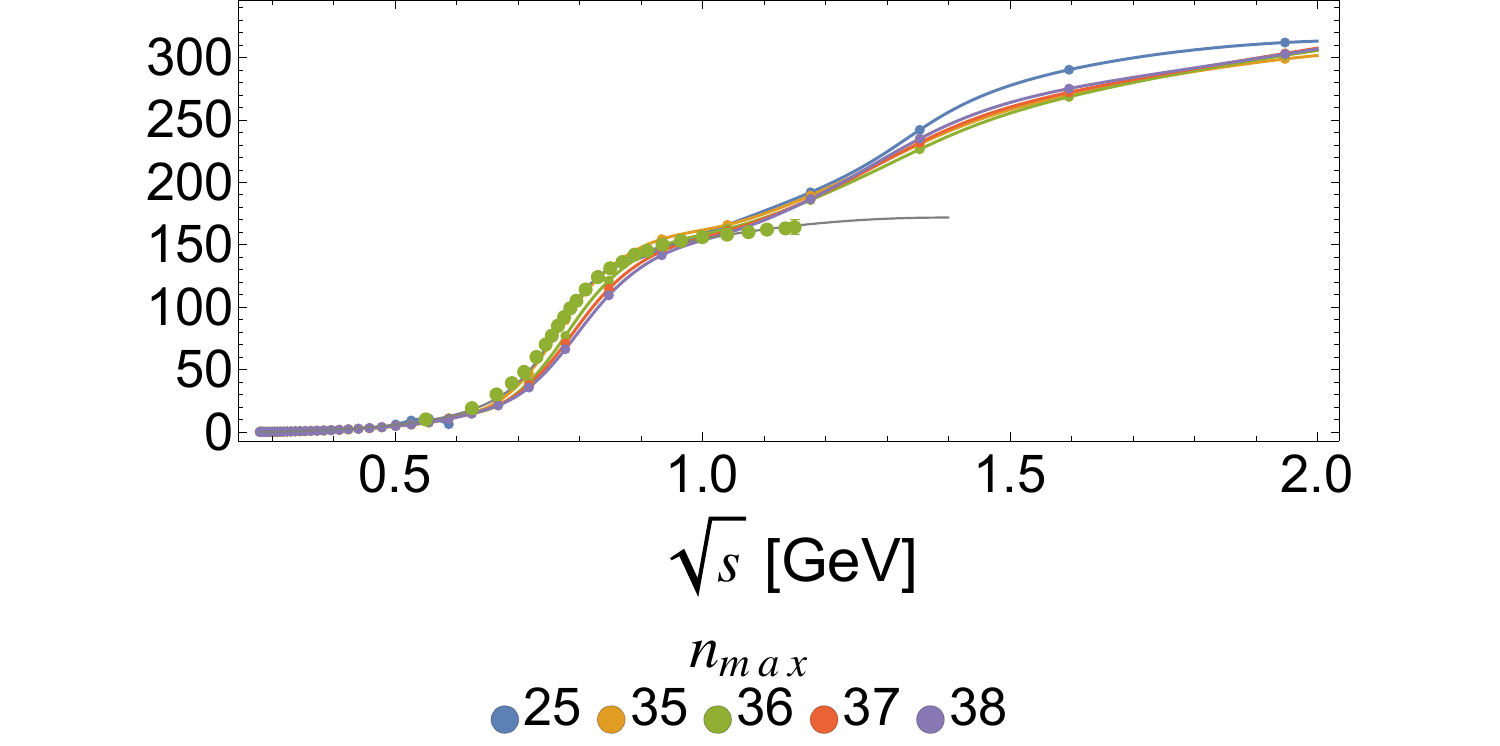}
    \put(15,30){\Large $\delta_1^1$}
\end{overpic}}
{\begin{overpic}[width=1.014\linewidth,trim={4.8cm 0.2cm 6.7cm 0cm},clip]{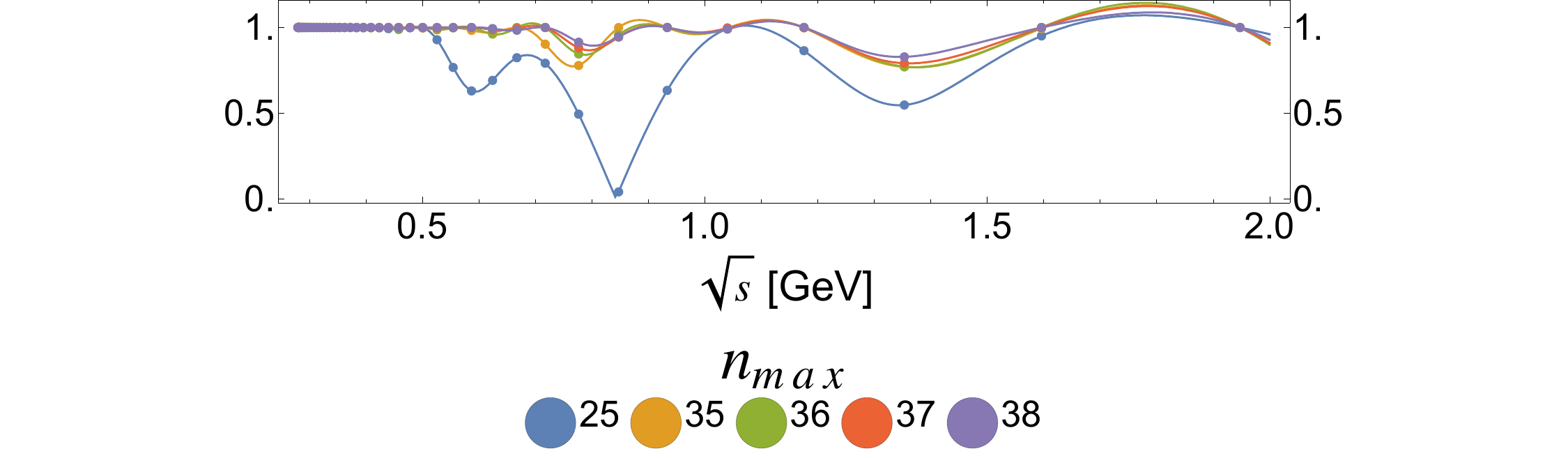}
\put(15,30){\Large $\eta_1^1$}
\end{overpic}}
\caption{$P1$ channel.}
\label{fig:cnmax}

\end{center}
\end{subfigure}
\caption{Convergence of phase shifts $\delta_\ell^I$ and elasticities $\eta_\ell^I$ for fixed $M = 49$ and $\ell_{\text{max}} = 11$. The gray curve shows the QCD phenomenological fit from \cite{Pelez}, where Kaon production has been removed. The green dots represent the experimental results of \cite{PhysRevD.7.1279}.
We also show the points (dots) at which unitarity was imposed (energies $s_i$ of Eq. \ref{energies}).
} 
\label{fig:SVZP1nmaxconvergence}
\end{figure}

 On Figure \ref{fig:SVZP1nmaxconvergence}   we further corroborate the $n_{max}$ dependence by examining  an amplitude found by imposing the full GTB constraints\footnote{This amplitude corresponds to the cyan point shown in Figure \ref{CSBtol}.}.   Figure \ref{fig:SVZP1nmaxconvergence} shows, in solid lines, the phase shifts $\delta_\ell^I$ and elasticities $\eta_\ell^I$ , defined by \begin{equation}
    S_\ell^I(s)=\eta(s)e^{2i \delta_\ell^I(s)},
    \label{phaseshifts}
\end{equation}  for the $S0$, $P1$, and $D0$ partial waves at fixed $\ell_{\text{max}} = 11$ and $M = 49$.   On top of the solid lines we plot added the points at which the GTB constraints were imposed, according to the maps of eq.s \eqref{energies} and \eqref{genM}.
Beyond $n_{\text{max}} \approx 29$, the amplitudes stabilize, with higher-order modes providing finer corrections.
At very high orders, however, the highly oscillatory character of the basis functions $\phi_n$ can interfere with numerical precision. The results show good agreement with experimental data (green dots) and phenomenological models (gray line) which we are going to discuss in the next section.

\subsection{Tolerance and Stability of GTB}
\label{app:tol}
\begin{figure}[H]
  \begin{center}    \includegraphics[width=0.8\textwidth,trim={0cm 6.1cm 0cm 2.8cm},clip]{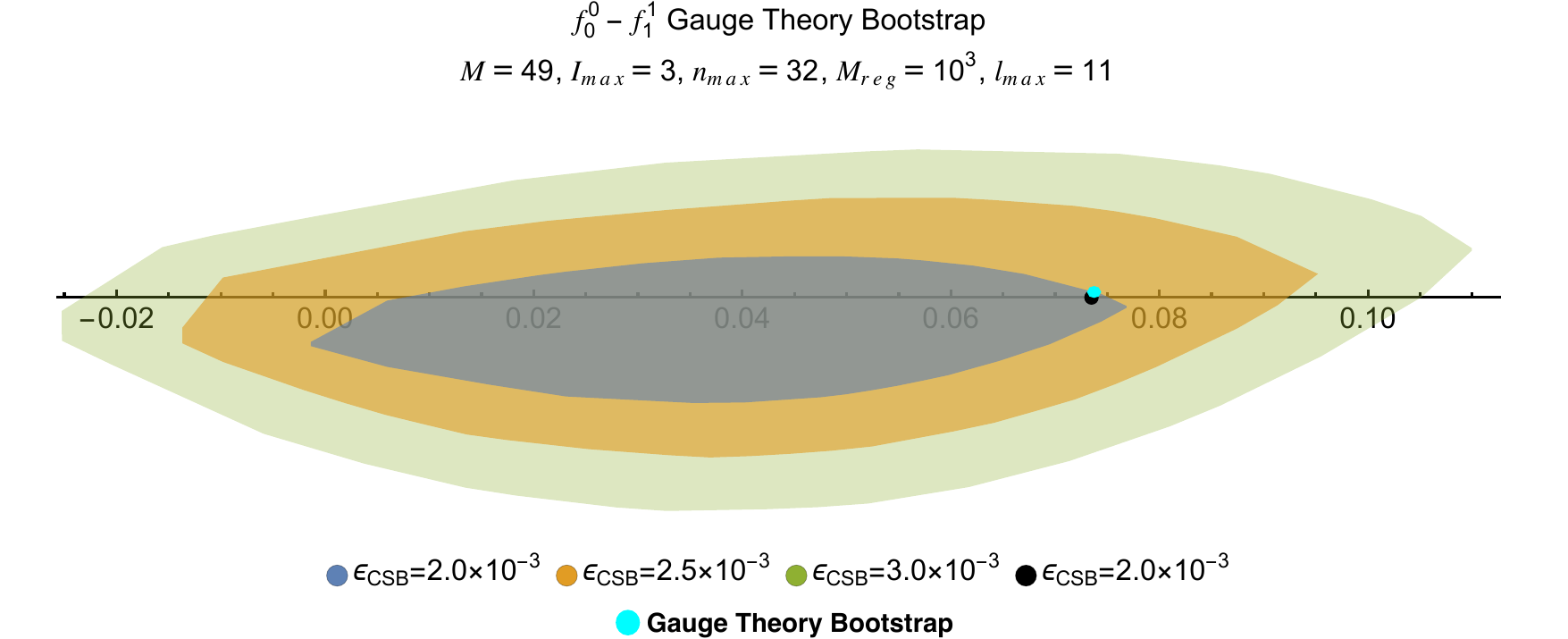}
  \end{center}
  \caption{\label{CSBtol} Gauge Theory Bootstrap $\varepsilon_{CSB}$ tolerance. The relation between $f_0^0(3)$ and $f_1^1(3)$, predicted by tree-level chiral perturbation theory, is taken as the $x$-axis. The black dot  corresponds to the physical point (i.e. $f_{\pi}=92MeV$) where tree-level chiral perturbation theory predicts the coordinates of $f_0^0(3)$ and $f_1^1(3)$. We highlight on cyan a point in the boundary of the allowed region for $\varepsilon_{CSB}=2.0\times 10^{-3}$. }
\end{figure}

Tolerances are chosen to be able to accommodate further  corrections of the UV and IR leading order predictions. Too strict tolerance could make the problem unfeasible as, for instance, higher order corrections are required to achieve unitarity on the chiral amplitude of eq. \eqref{ampchi}. Varying $\varepsilon_{CSB}$ while fixing the other tolerances is illustrated in Figure \ref{CSBtol}. There, the relation between $f_0^0(3)$ and $f_1^1(3)$, predicted by chiral perturbation theory, i.e. eq. \eqref{relationcsb}, is taken as the $x$-axis. The black dot  corresponds to the physical value where chiral perturbation theory predicts the coordinates of $f_0^0(3)$ and $f_1^1(3)$ through eq. \eqref{eq:fchiral} at the experimentally-observed pion decay constant $f_{\pi}=92MeV$. We highlight on cyan the closest boundary point --for $\varepsilon_{CSB}=2.0\times 10^{-3}$-- to the physical point. The phase shifts of the  bootstrapped S-Matrix elements associated to the cyan point are shown in the next Section in Figure \ref{fig:SVZP1nmaxconvergenc3}.

\section{Bootstrapped $S-$Matrix}

\label{sec:res}
The numerical convergence analysis of Section \ref{sec:dependences} allows us to select a set of parameters that stabilizes the bootstrap and optimizes computational performance. Being conservative, we choose 
$M=49$, $n_{max}=37$ and $l_{max}=11$ 
which reduces the runtime of the optimization problem from roughly 30 minutes in \cite{He:2024nwd} to about one minute on a standard computer.

For the tolerance parameters $\varepsilon_{CSB}, \varepsilon_{SVZ}$, and $\varepsilon_{FF}$ of the IR and UV constraints of  eq.s \eqref{CSBconst}, \eqref{SVZconst} and \eqref{FF}, respectively, we keep the same order of magnitude as in \cite{He:2024nwd}. We take then:
\vspace{-0.44cm}\begin{equation}
  \varepsilon_{CSB}=2\times 10^{-3},\quad  \vec{ \varepsilon}_{SVZ}= \left(\begin{matrix}
    1\times 10^{-7}\\
    4\times 10^{-6}\\
    4\times 10^{-6}
  \end{matrix}\right) ,\quad \vec{\varepsilon}_{FF}= \left(\begin{matrix}
    3\times 10^{-8}\\
    2\times 10^{-6}\\
    2\times 10^{-6}
  \end{matrix}\right), \quad M_{reg}=10^3.
  \label{parGTB}
\end{equation}
Here, each element of the vectors $\vec{\varepsilon}_{FF}$ and $\vec{\varepsilon}_{SVZ}$ are the tolerance associated to the  constraints  for the $S0$, $P1$, and $D0$ partial waves, respectively. We consider the resulting bootstrapped amplitude associated to the cyan point in Figure \ref{CSBtol}. This is the closest solution --on the boundary of the GTB-allowed region--to the physical value of $f_\pi = 92\text{MeV}$, here showed as a black dot. 

\subsection{Partial Waves}
We present our bootstrapped  $S$-matrix (cyan point in Fig. \ref{CSBtol}) in terms of the phase shifts $\delta_\ell^I$ and elasticities $\eta_\ell^I$, defined in eq.~\eqref{phaseshifts}, and compare it with experimental data~\cite{PhysRevD.7.1279}. Since our setup consider only two quark flavors, we also compare it with a phenomenological QCD model \cite{Pelez} where Kaon production is removed (as there is no strange quark producing it). 

\begin{figure}[H]
\centering
\begin{subfigure}[b]{0.49\linewidth}
\begin{overpic}
   [width=1.007\linewidth,trim={2.5cm 4.65cm 2.5cm 0cm},clip]{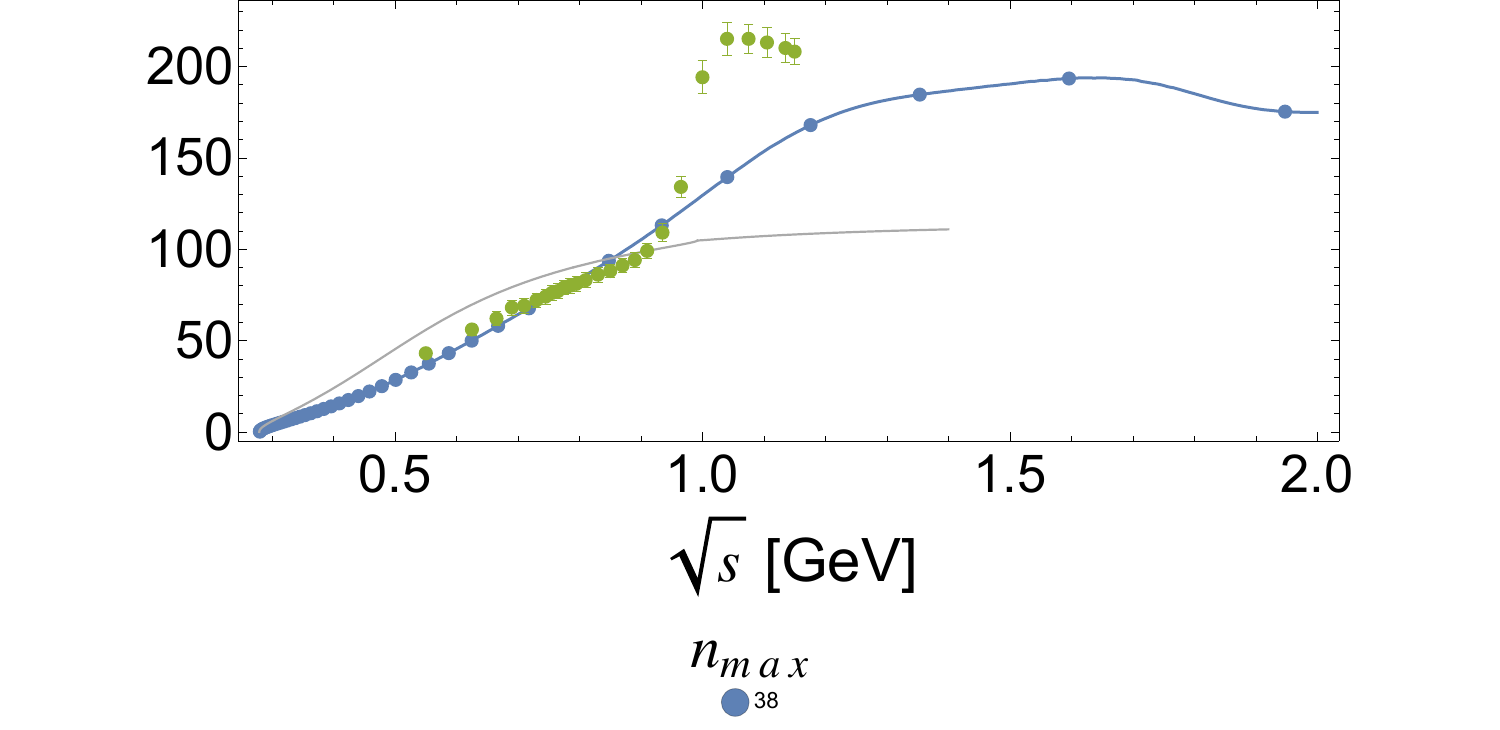}\put(15,20){\Large $\delta_0^0$} 
\end{overpic}
\begin{overpic}[width=1\linewidth,trim={5.5cm 3.6cm 7.3cm 0cm},clip]{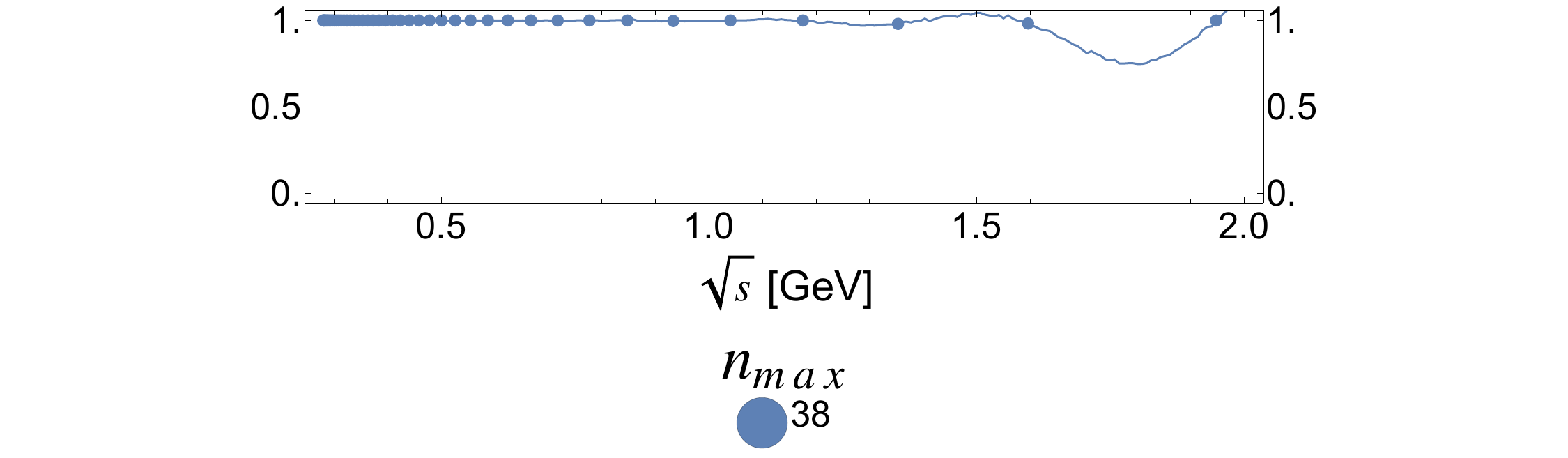}\put(15,17){\Large $\eta_0^0$}\end{overpic}
\caption{ $S0$ channel.}
\label{fig:cnmaxd}
\end{subfigure}
\begin{subfigure}[b]{0.49\linewidth}
\begin{overpic}[width=1.007\linewidth,trim={2.5cm 5cm 2.5cm 0cm},clip]{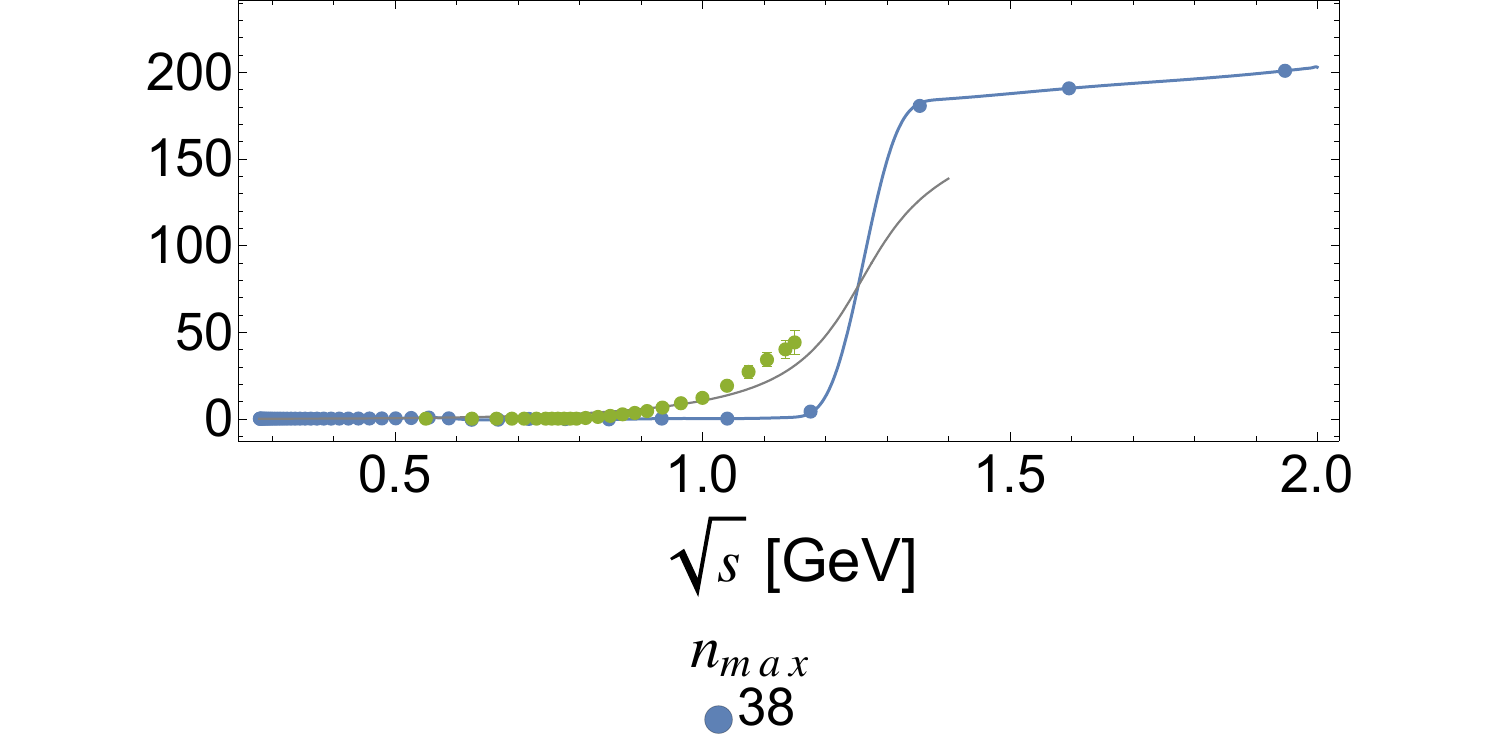}\put(15,20){\Large $\delta_2^0$}\end{overpic}
\begin{overpic}[width=1\linewidth,trim={5.5cm 3.6cm 7.3cm 0cm},clip]{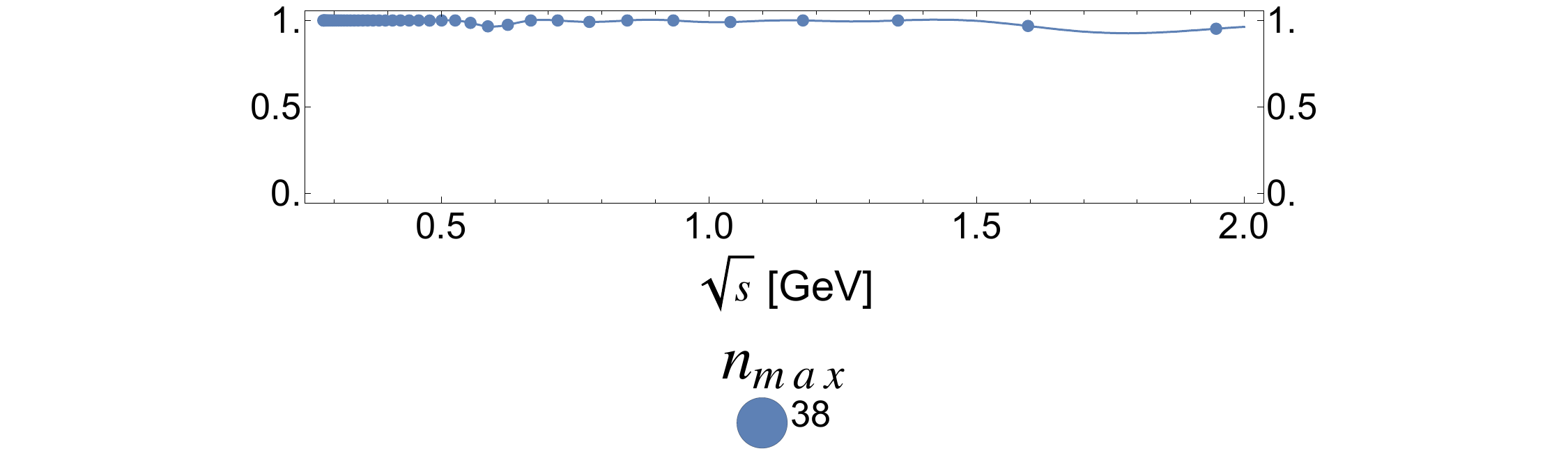}\put(15,17){\Large $\eta_2^0$}\end{overpic}
\caption{ $D0$ channel.}
\label{fig:dnmaxd}
\end{subfigure}

\begin{subfigure}[b]{0.49\linewidth}
\begin{overpic}[width=1.007\linewidth,trim={2.5cm 4.65cm 2.54cm 0cm},clip]{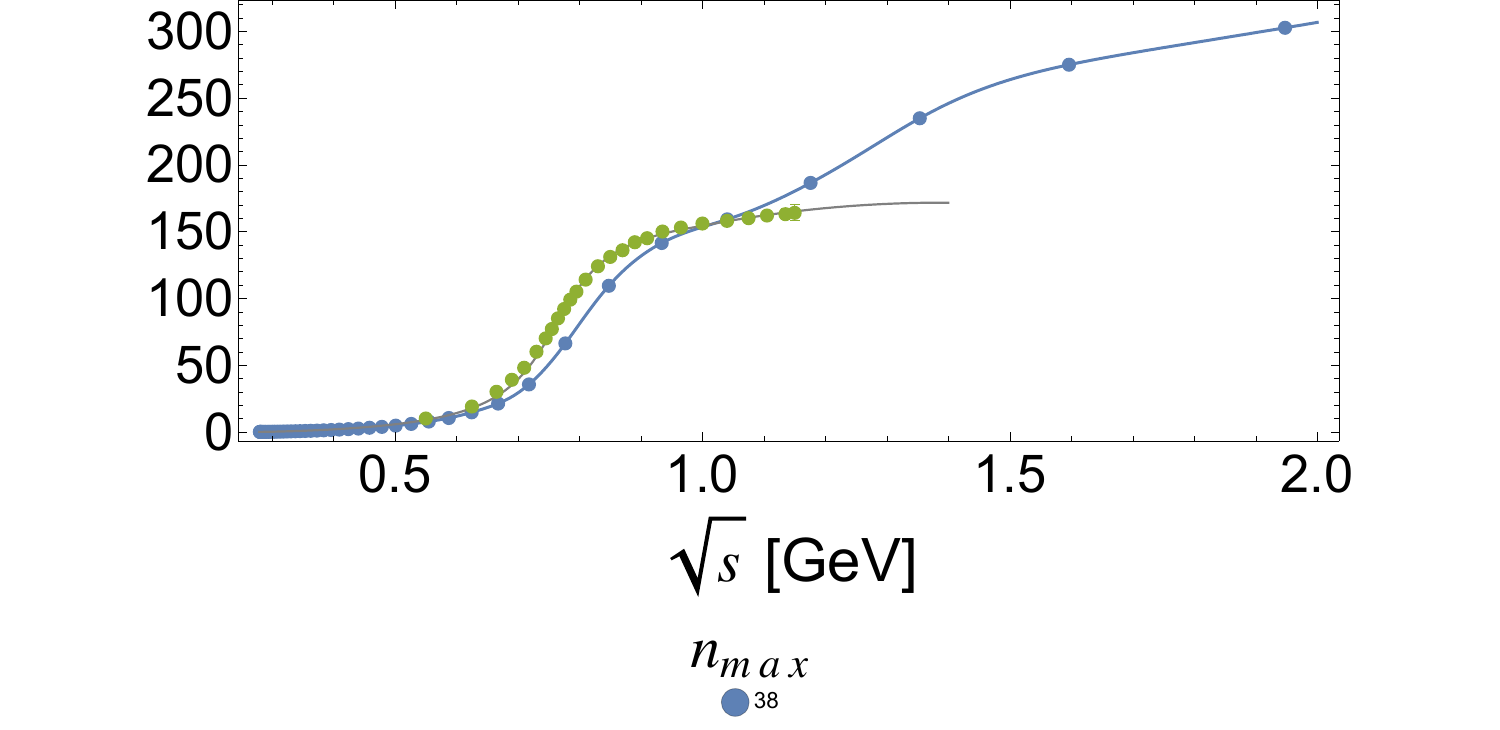}\put(15,20){\Large $\delta_1^1$}\end{overpic}
\begin{overpic}[width=1\linewidth,trim={4.8cm 3.6cm 6.7cm 0cm},clip]{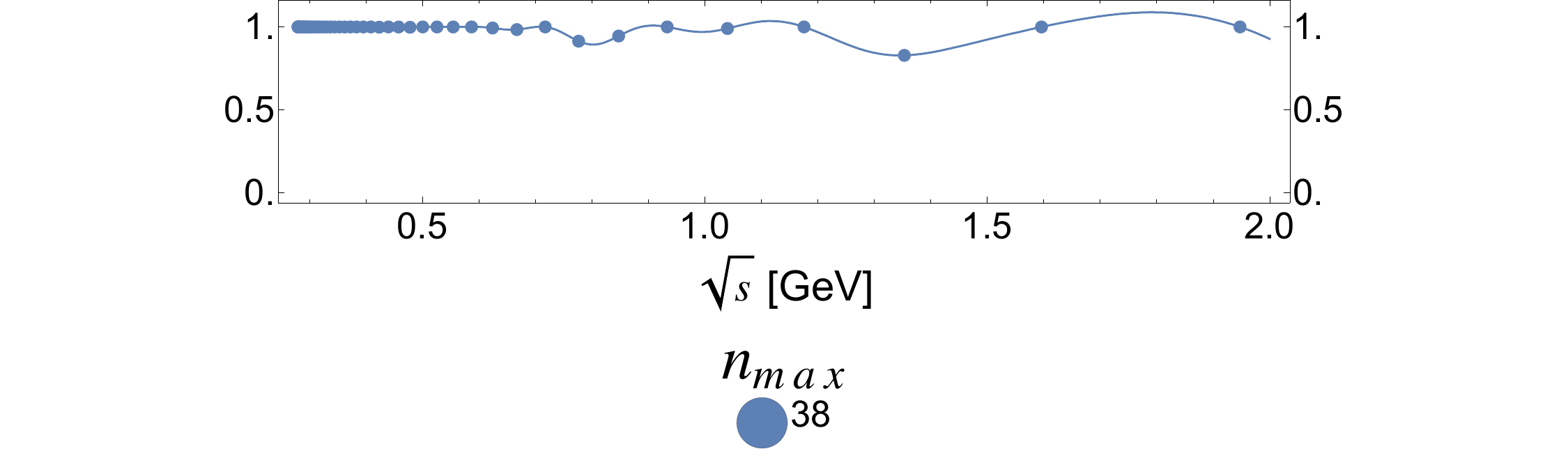}\put(15,17){\Large $\eta_1^1$}\end{overpic}
\caption{ $P1$ channel.}
\label{fig:anmaxd}
\end{subfigure}
\begin{subfigure}[b]{0.49\linewidth}
\begin{overpic}[width=1.007\linewidth,trim={1cm 4.65cm 4.53cm 0cm},clip]{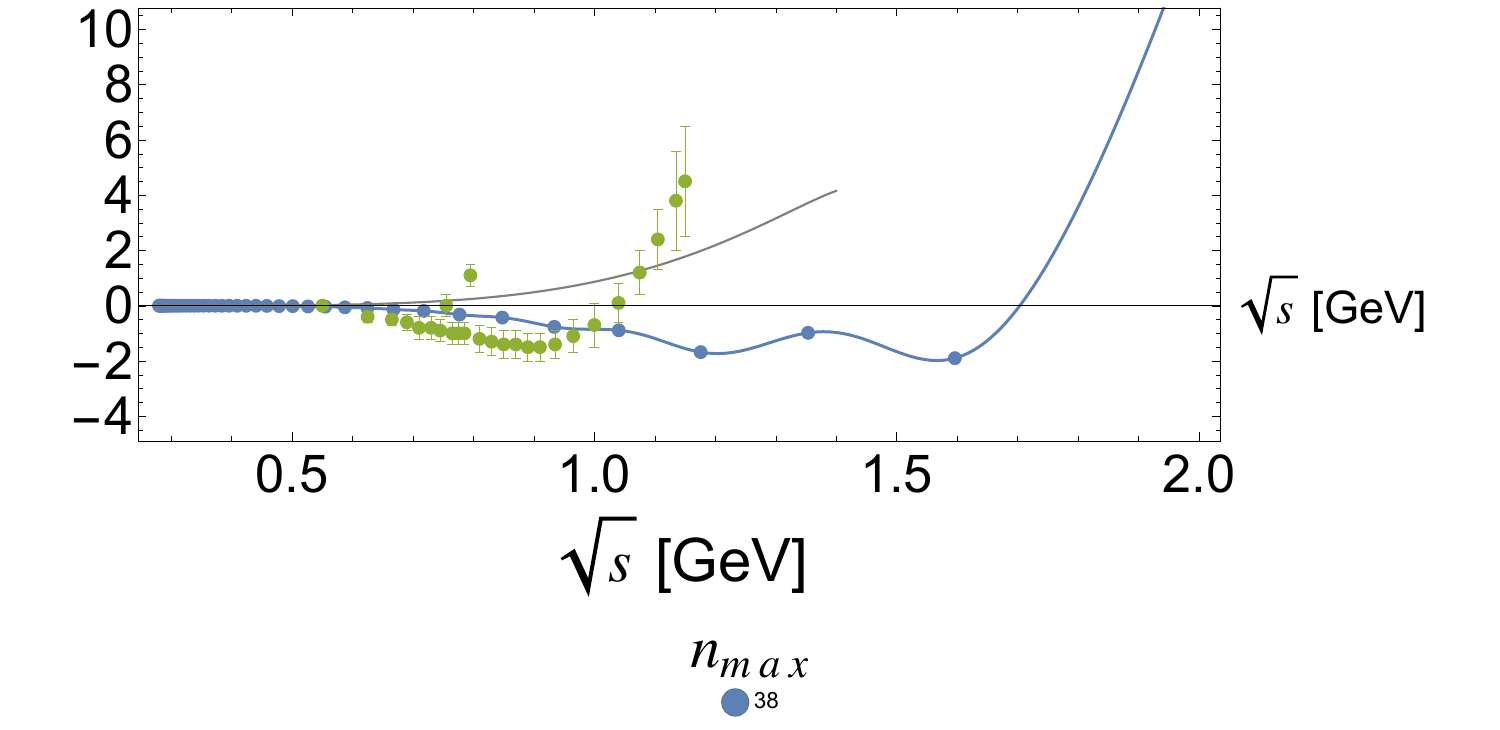}\put(15,20){\Large $\delta_3^1$}\end{overpic}
\begin{overpic}[width=1\linewidth,trim={4.9cm 3.6cm 6.7cm 0cm},clip]{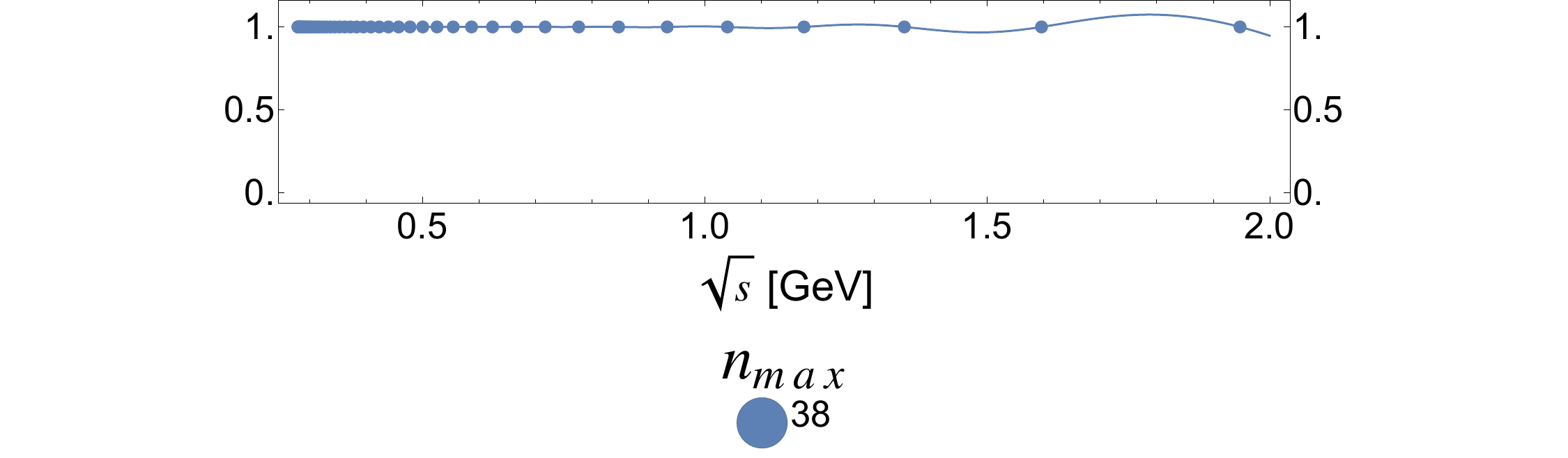}\put(15,17){\Large $\eta_3^1$}\end{overpic}
\caption{ $F1$ channel.}
\label{fig:bnmaxd}
\end{subfigure}

\begin{subfigure}[b]{0.49\linewidth}
\begin{overpic}[width=1.007\linewidth,trim={2.5cm 4.65cm 2.61cm 0cm},clip]{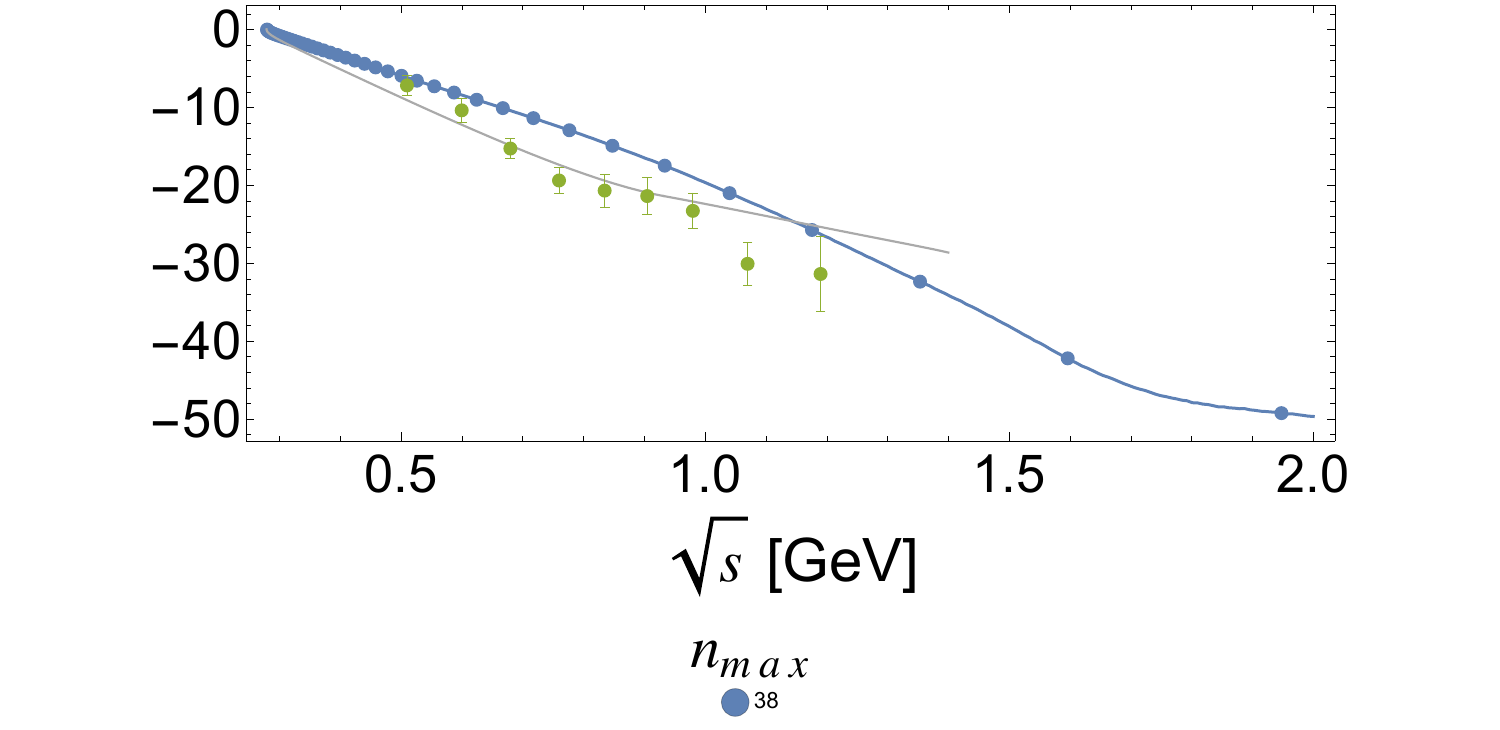}\put(15,20){\Large $\delta_0^2$}\end{overpic}
\begin{overpic}[width=1\linewidth,trim={4.6cm 3.6cm 6.7cm 0cm},clip]{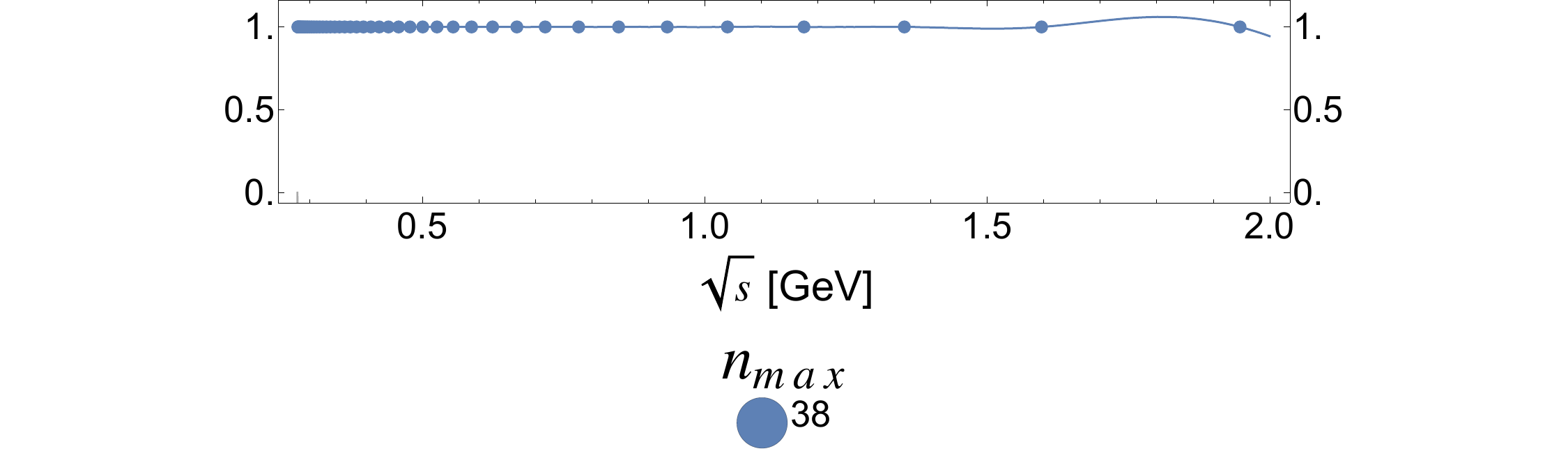}\put(15,17){\Large $\eta_0^2$}\end{overpic}
\caption{ $S2$ channel.}
\label{fig:enmaxd}
\end{subfigure}
\begin{subfigure}[b]{0.49\linewidth}
\begin{overpic}[width=1.02\linewidth,trim={2.4cm 4.65cm 2.5cm 0cm},clip]{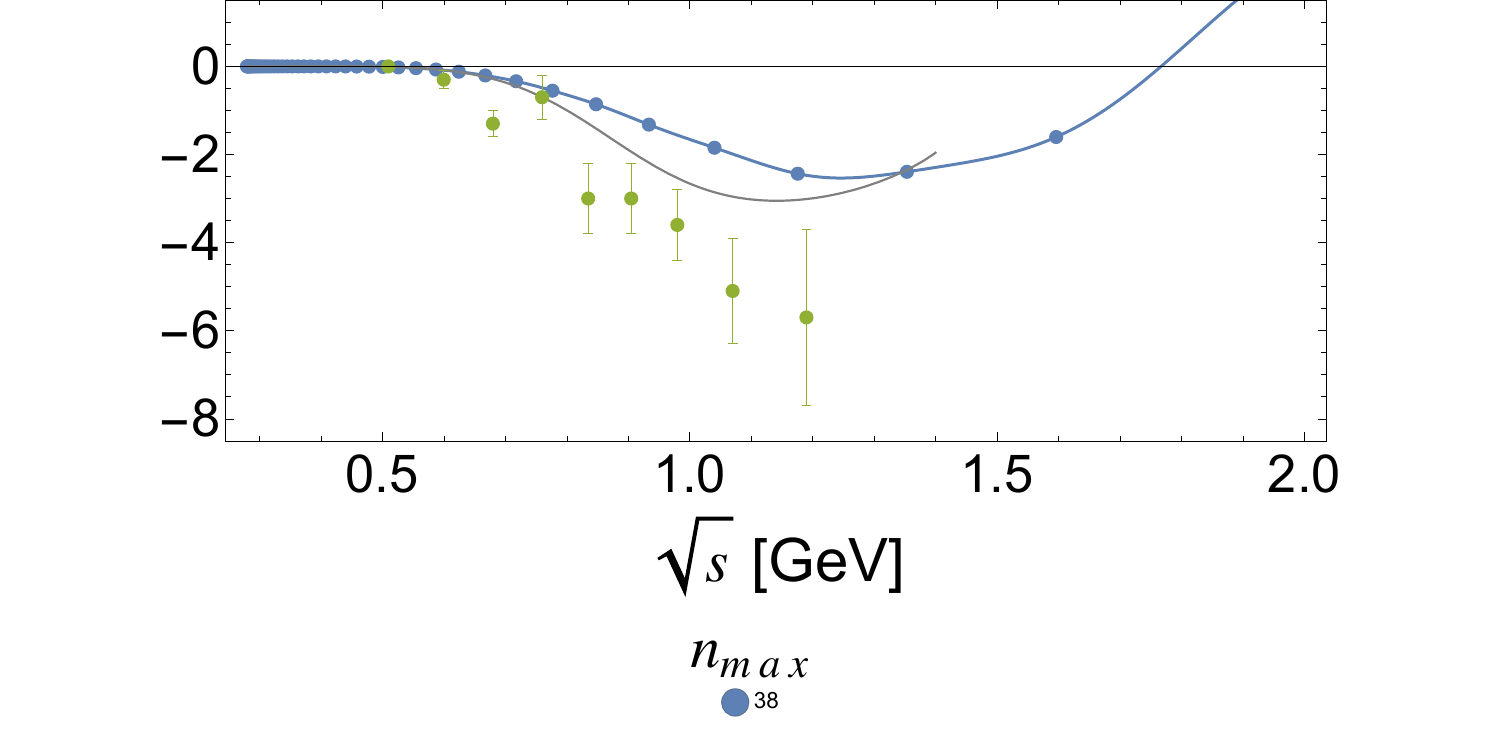}\put(15,20){\Large $\delta_2^2$}\end{overpic}
\begin{overpic}[width=1\linewidth,trim={4.9cm 3.6cm 6.7cm 0cm},clip]{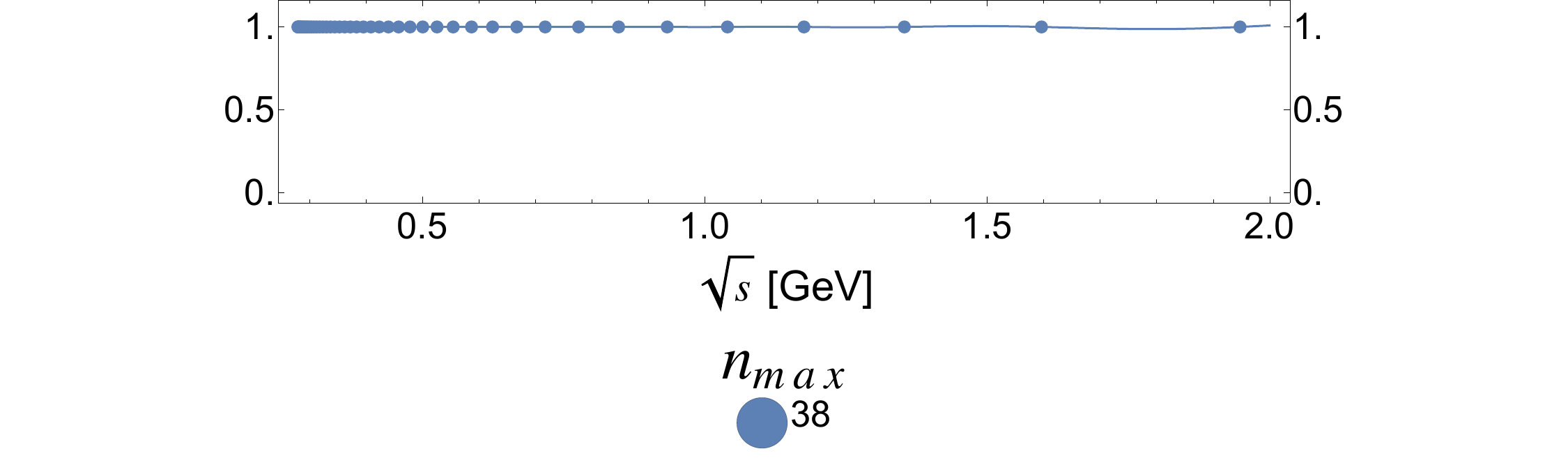}\put(15,17){\Large $\eta_2^2$}\end{overpic}
\caption{ $D2$ channel.}
\label{fig:fnmaxd}
\end{subfigure}

\caption{Bootstrapped phase shifts $\delta_\ell^I$ and elasticities $\eta_\ell^I$ for the first six partial waves. 
 The gray curve  shows the two-quark QCD phenomenological fit from \cite{Pelez} and the green dots represent the experimental results of \cite{PhysRevD.7.1279}, where Kaon production has been removed. The continuous blue line represents our GTB result. For reference, we also have added the points (blue dots) at which unitarity was imposed (energies $s_i$ of Eq. \ref{energies}).
}
\label{fig:SVZP1nmaxconvergenc3}
\end{figure}

Figure \ref{fig:SVZP1nmaxconvergenc3} displays the first six phase shifts for our bootstrapped $S-$matrix. We observe that below $1GeV$ both data shows good agreement with the bootstrapped amplitude. Above $1GeV$ the $S0$ channel shows a phase rise due to kaon production in the experimental data—a feature not expected in our setup since the strange quark is excluded. The phenomenological model, with kaon contributions subtracted, aligns better, after $1 GeV$, with our bootstrapped phase shifts. All together we observe good agreement with the literature checking the consistency of the GTB results of \cite{He:2024nwd,He:2023lyy}.

From these figures, we can extract the resonances of the scattering. For example, Figure \ref{fig:cnmaxd} shows a broad rise near $\sqrt{s} \sim 200–800 MeV$ suggesting an identification with the $\sigma$ meson. Similarly, Figure \ref{fig:anmaxd} shows a rise around $\sqrt{s} \sim 800$ MeV indicating the $\rho(770)$ meson. In the $D0$ channel (Figure \ref{fig:enmaxd}), we observe the $f_2(1270)$ resonance.

\subsection{Resonances and Pole Structure}
\begin{figure}[H]
\raisebox{-1.4cm}{
\begin{subfigure}{0.315\linewidth}
\begin{overpic}[width=1.1\linewidth,trim={1.39cm 2.6cm 1.1cm 1.9cm},clip]{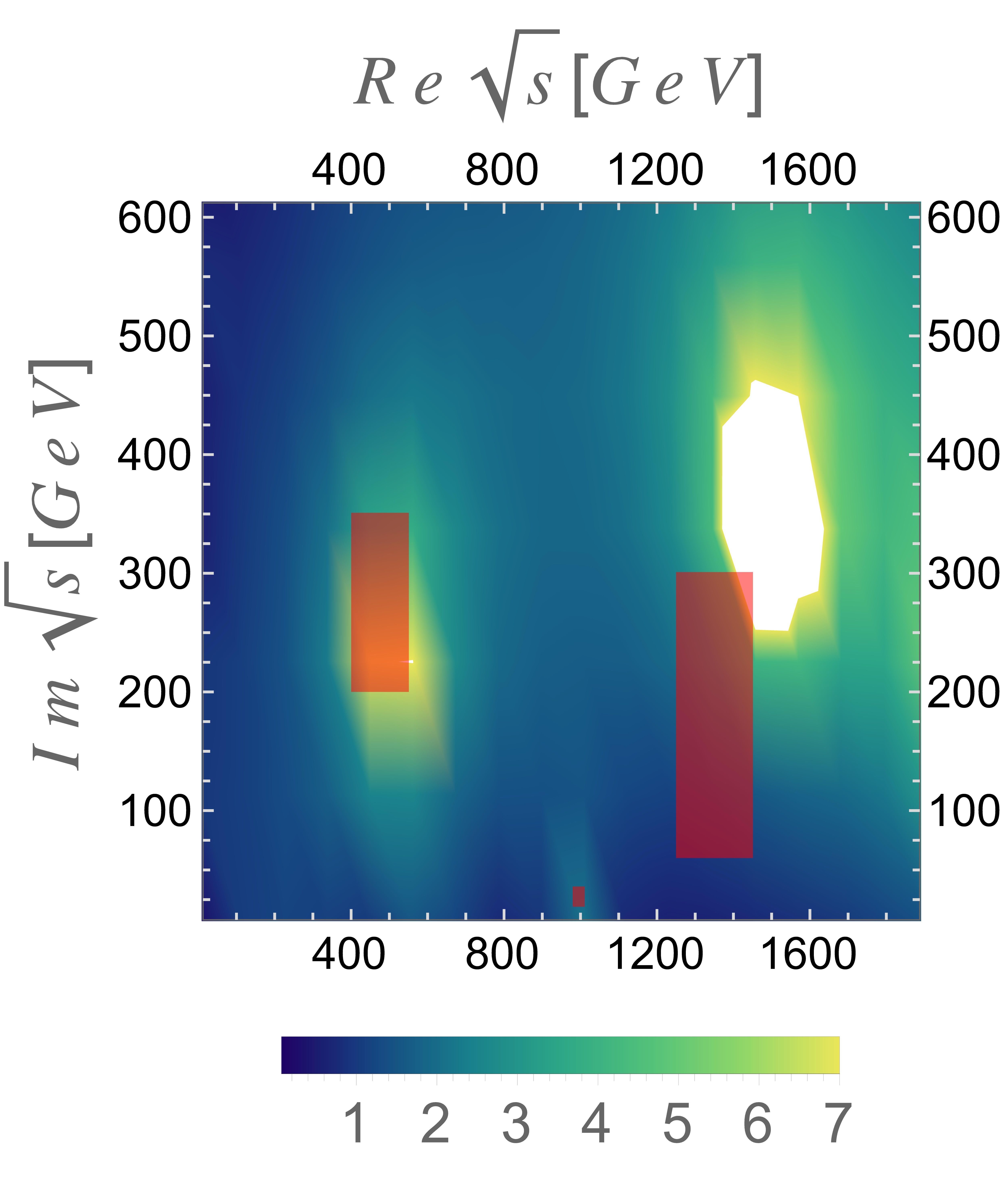}
\put(26,58){\Large \color{red}$\sigma$}
\put(47,14){\Large \color{red}$f_0$}
\put(60,52){\Large \color{red}$f_0'$}
\end{overpic}
\vspace{0.83cm}
\caption{$|S_0^0(z)|$\label{fig:Polessharpnessgrid1}}
\end{subfigure}}
\hspace*{0.4cm}
\raisebox{-1.25cm}{\begin{subfigure}{0.32\linewidth}

\begin{overpic}[width=0.96\linewidth,trim={2.52cm 0.4cm 1.1cm 1.91cm},clip]{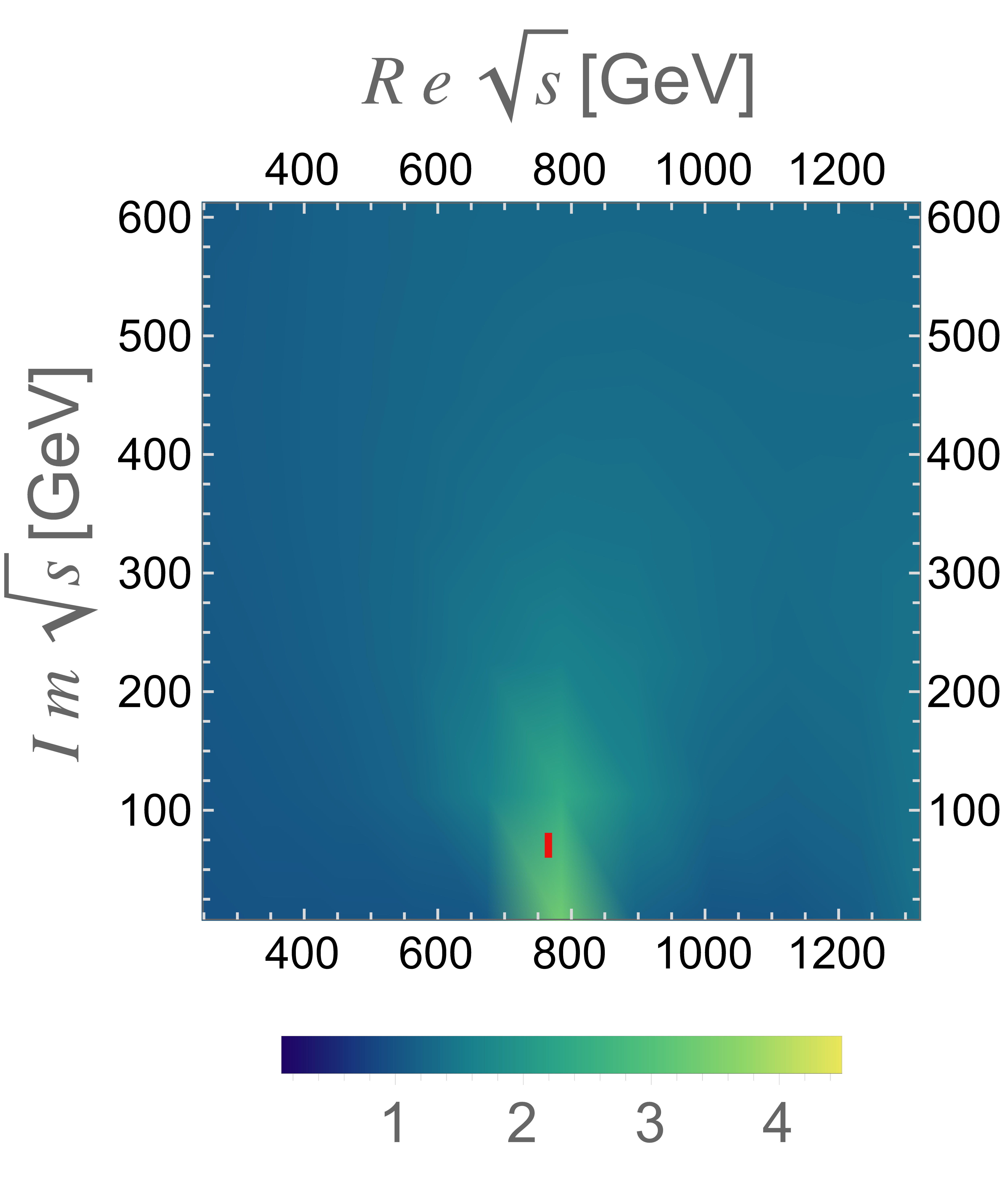}
\put(30,35){\Large \color{red}$\rho$}
\end{overpic}
\caption{$|S_1^1(z)|$\label{fig:Polessharpnessgrid2}}
\end{subfigure}}
\raisebox{-1.4cm}{\hspace*{-0.2cm}\begin{subfigure}{0.32\linewidth}
\begin{overpic}[width=1.09\linewidth,,trim={2.52cm 2.6cm 0cm 1.9cm},clip]{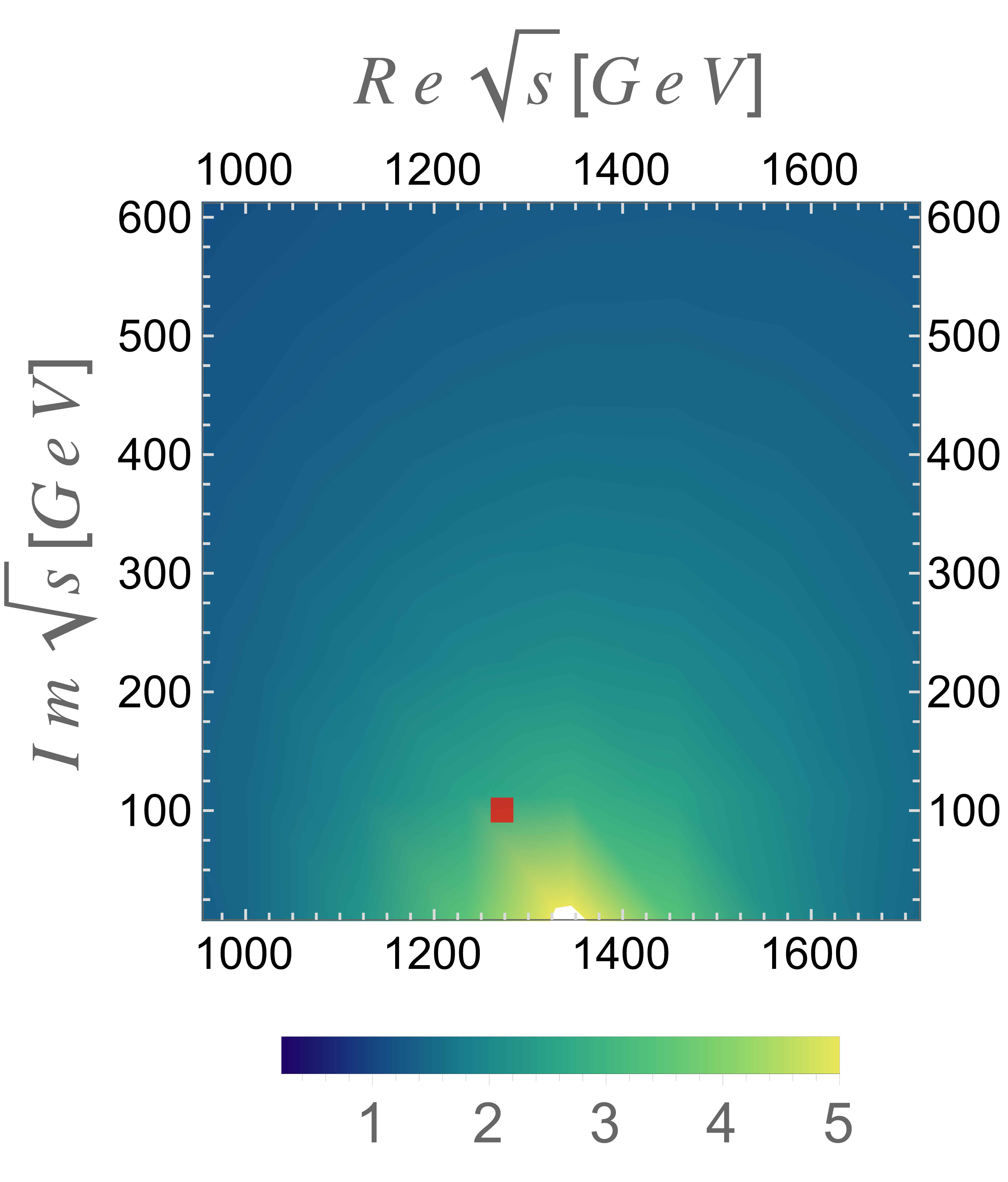}
\put(28,25){\Large \color{red}$f_2$}
\end{overpic}
\vspace{0.83cm}
\caption{$|S_2^0(z)|$\label{fig:Polessharpnessgrid3}}
\end{subfigure}}

\caption{$S$-matrix element partial waves poles on the second complex sheet. The $x-$axis and $y-$axis represents $Re[\sqrt{s}]$ and  $Im[\sqrt{s}]$, in GeV, respectively. Red regions  indicate PDG-reported meson pole ranges \cite{Workman:2022ynf}.}
\label{fig:Polessharpnessgrid}
\end{figure}

Since the bootstrapped S-matrix can be directly evaluated across the first and second complex sheet, we plot the poles of the $S0$, $P1$ and $D0$ channels,  evaluated at the second complex sheet, in Fig. \ref{fig:Polessharpnessgrid}. Here we have colored in red the regions where hadrons were experimentally identified, according to the Particle Data Group (PDG) \cite{Workman:2022ynf}.

Figure \ref{fig:Polessharpnessgrid1} shows the presence of the $\sigma$ meson (a.k.a the $f_0(550)$ resonance), a second pole around $\sqrt{s}\sim (1000+i \cdot 30)MeV$ and a third pole around $\sqrt{s}\sim (1400+i \cdot 300)MeV$.  
In the $P1$ channel, Figure \ref{fig:Polessharpnessgrid2}, we identify the $\rho(770)$ with $\sim 7\%$ error relative to the PDG value. For the $D0$ channel, Figure \ref{fig:Polessharpnessgrid3}, we identify the $f_2(1270)$  resonance.
The poles located far from the real axis --such as the $\sigma$ meson-- were  difficult to identify using the interpolation-based methods of \cite{He:2024nwd}. With our implementation they can be easily found by evaluating the explicit amplitude across the second complex sheet. Doubling the resolution of the evaluation grid significantly enhances the precision of the pole location. From these we deduce the T-Matrix Poles shown in Table \ref{tab:poles}.

\begin{table}[H]
\centering

\begin{tabular}{|c|c|c|c|c|}
\hline
\textbf{State} & \multicolumn{2}{c|}{\textbf{PDG best value \cite{Workman:2022ynf}}} & \multicolumn{2}{c|}{\textbf{Bootstrap}} \\
\cline{2-5}
 & Re$[\sqrt{s}]$ [MeV] & Im$[\sqrt{s}]$ [MeV] & Re$[\sqrt{s}]$ [MeV] & Im$[\sqrt{s}]$ [MeV] \\
\hline
$\sigma$ & $400$--$550$ & $200$--$350$ & $484.375$ & $250$ \\
$\rho$ & $761$--$765$ & $71$--$74$ & $764$ & $31.25$ \\
$f_2(1270)$ & $1260$--$1283$ & $90$--$110$ & $1350$ & $10$ \\
\hline
\end{tabular}

 \caption{\label{tab:poles}T-Matrix Poles for the first three partial waves.}
\end{table}

\section{Conclusions}
\label{sec:out}
In this work, we have implemented a discrete basis parametrization of the Gauge Theory Bootstrap \cite{He:2023lyy,He:2024nwd}, to the case of QCD with $N_f = 2$, for the $2$-to-$2$ pion scattering. 
The results obtained are aligned from those of the original implementation \cite{He:2023lyy,He:2024nwd}, exhibiting good agreement with  both phenomenological \cite{Pelez} and experimental data \cite{PhysRevD.7.1279}. In particular,  the parametrization hereby introduced of the GTB allowed us the identification of  the $S$-matrix's resonances  associated to the $S0$, $P1$, and $D0$ partial waves. Our analysis reveals poles corresponding to the $\sigma$, $\rho(770)$, and $f_2(1270)$ mesons, in good agreement with the reported values of the PDG \cite{Workman:2022ynf}. Since the discrete basis give the parametrization of the S-matrix in all it's partial waves, in principle, other channels could be explored, although we haven't introduced  their explicit UV information and so we do not expect enlightening  information from those bootstrapped partial waves. It would be interesting to explore --and quantify-- how these additional channels may modify the pole structure of the dominant, found poles.

The key advantages of the discrete basis parametrization is its computational efficiency. Compared to the original implementation in \cite{He:2024nwd}, which required approximately 30 minutes per run, the new parametrization reduces the runtime to about one minute on a standard computer. 
This improvement enabled us to perform checks of the consistency and convergence of the bootstrap procedure and opens the door for a more exhaustive exploration of the GTB framework. In particular, it will allows us to investigate the uniqueness of the solutions, as suggested in \cite{he2025gaugetheorybootstrappredicting}. 

 Overall, the newly obtained results further corroborate the consistency of the Gauge Theory Bootstrap and its power for studying the 
 strongly coupled physics of gauge theories.

\nocite{*}
\section*{Acknowledgments}
We would like to thank Yifei He for useful discussions and comments on the draft. 

\bibliographystyle{JHEP}


\phantomsection 
\pdfbookmark[1]{References}{References} 

\bibliography{biblio}

\end{document}